\documentclass[a4paper]{article}
\usepackage{graphicx}
\usepackage{amsmath}
\usepackage{authblk} 
\usepackage{siunitx} 
\usepackage[numbers]{natbib}  
\usepackage{hyperref}
\usepackage{xcolor}
\usepackage{geometry}
\usepackage{amssymb}
\usepackage{multirow}
\usepackage{booktabs}
\usepackage{makecell}
\usepackage{url}

\usepackage{adjustbox}

\usepackage[symbol]{footmisc}

\title{A novel 3D food printing technique: achieving tunable porosity and fracture properties via liquid rope coiling}    

\author[1]{Aref Ghorbani$^*$}
\author[1]{Sophia Jennie Giancoli$^*$}
\author[2]{Seyed Ali Ghoreishy}
\author[2]{Martijn Noort}
\author[1]{Mehdi Habibi}

\affil[1]{Laboratory of Physics and Physical Chemistry of Foods, Wageningen University \& research, 6708 WG Wageningen, The Netherlands}
\affil[2]{Fresh Food \& Chains, Wageningen University \& research, 6708 WG Wageningen, The Netherlands}

\hypersetup{
	colorlinks=true,
	linkcolor=blue,
	filecolor=blue,      
	urlcolor=blue,
	citecolor=blue,
	pdftitle={Preprint; Coiling 3D Food Printing},
	pdfauthor={Aref Ghorbani and Sophie Jennie Giancoli},
	pdfkeywords={Additive manufacturing, 3D food printing, Liquid rope coiling, Texture, Texture analysis, X-ray computed tomography, Rheology, Food fracture},
}

\newcommand{\keywords}[1]{%
	\textbf{Keywords:} #1
}
         
\begin{document}
    \maketitle
    \footnotetext[1]{These authors contributed equally to this work.}
    
    \begin{abstract}
           We present a 3D food printing (3DFP) method to create coiled structures, harnessing the liquid rope coiling effect as a rapid method of food printing with tunable fractural properties. By studying the printability and coil-forming ability of pea, carrot, and cookie dough inks, we identified optimal printing parameters to induce steady and controlled coiling, enabling the creation of coiled structures with tunable porosities using a single nozzle. Fracture profiles from post-processed coiled structures showed complex responses but presented direct correlations between the porosity and textural parameters, including hardness, brittleness, and initial stiffness. This study provides a foundation for the fabrication of coiled food structures using 3DFP and highlights its potential application in designing textural properties and a range of unique sensory experiences.  
    \end{abstract}
    

\section{Introduction}

3-Dimensional food printing (3DFP) is a rapidly advancing field, receiving increasing interest from both academia and the food industry. Extrusion-based printing is the most used method in 3DFP, where a food product is based on a digital design, printed layer by layer until the full design is produced \cite{dankar2018, gholamipour2020}. 3DFP provides new food processing opportunities that are unattainable through conventional food processing methods, which facilitates the creation of personalized products that suit consumers’ taste, textural, and nutritional needs \cite{sun2015}. This can be especially useful for consumers with specific needs like dysphagia patients, athletes, and children, or for consumers who want to try new organoleptic experiences such as pieces of chocolate breaking apart in unconventional ways \cite{dankar2018, godoi2016, lorenz2022, Souto2022}. 

Fluid instabilities that are intrinsic to the printing process are often among the major challenges involved in 3DFP and result in lower printing resolution or structural inconsistencies \cite{zhang2022}. These instabilities can occur for a myriad of reasons, but they are often due to inadequate configuration of the printer or inconsistencies in the filament. One such instability stems from a well-studied phenomenon known as liquid rope coiling (LRC), which occurs when the nozzle height is larger than a critical value, leading to the coiling of the falling stream of viscous liquid upon hitting the substrate \cite{Habibi2010, Ribe2006a}. In daily life, LRC is typically observed when pouring honey on toast \cite{barnes1958}, and has been thoroughly investigated in recent decades \cite{brun2015, Maleki2004, ribe2004, Ribe2006, Ribe2006a, ribe2012, Habibi2010, Habibi2011, Habibi2006, Habibi2008, Habibi2014}.

The nozzle height is usually minimized to avoid disruptive effects originating from LRC in extrusion-based 3D printing. Yet, despite its often undesired nature in 3D printing, LRC can be exploited as a tool for structural designing. In this manner, it has been used to create striking geometric structures across various materials and scales such as coiled molten glass and ceramics \cite{brun2017, mansoori2018}, nanofiber pottery sculptures \cite{Kim2010}, microfluidic channels \cite{yang2017}, and variable stiffness foams \cite{lipton2016}. Recently, LRC was used to create elastomer structures with graded porosity and distinct mechanical properties for actuation in soft robotics \cite{Willemstein2022, Willemstein2024}. Given the potential of utilizing LRC to produce diverse structures with tunable porosity, we aim to leverage LRC in a 3DFP setup to establish a framework for creating food structures with novel geometries and textural properties in a relatively rapid manner.

In this study, we exploited LRC in 3DFP to create coiled food structures using different food-based inks (carrot paste, pea paste, and cookie dough). Different printing conditions, including deposition height and extrusion flow rate, were investigated to identify the optimal coiling conditions, which were used to create coiled structures with varying degrees of porosity. The porosity was shown to be relevant for programming the mechanical properties of the coiled structures. This study shows that LRC-3DFP is an innovative food structuring method that holds promise for the manufacturing of products with unique textural and sensorial properties and prompts further research in this field.  

\section{Materials and Methods}

\subsection{Preparation of 3DFP inks}
Carrot and pea 3DFP inks were prepared by mixing carrot powder obtained from residue pulp after juice extraction and toasted pea flour from milled wrinkled green peas with demineralized water in ratios of 12.77:87.23 and 50:50, respectively; see the Supplementary Information (SI) A.1 for more details. These powder-to-water ratios were pre-determined to provide pastes with desirable extrudability for 3DFP. Mixing was performed in a speedmixer (DAC 150.1 FV-K, FlackTek) at 3500 rpm for 2 minutes in 185 mL PP100 cups. Furthermore, printable cookie dough was prepared based on methods described in \cite{piovesan2020}, further elaborated upon in SI A.1. 

\subsection{3D printing and characterization of coiled structures}
Extrusion printing was performed using the ‘SUPREME’ 3D food printer, which was built in-house and developed by TNO (The Netherlands). This printer consists of a syringe-based extruder powered by a stepper extruder motor (maximum driving force of 1400 N), a temperature-controllable syringe holder for 30 mL plastic syringes, a piston plunger, and a force sensor. This setup enabled us to vary the nozzle and syringe temperature (\(T_{\text{nozzle}}, T_{\text{syringe}}\) [°C]), printing height (H [mm]), printing (travel) speed (\(U_{xy}\) [mm/s]), piston speed (U [mm/s]), and the volumetric flow rate (Q [\(\text{mm}^3/\text{s}\)]). The latter two parameters were calculated as described in SI A.2. LRC was controlled by varying H, \(U_{xy}\) and Q while maintaining a constant \(T_{\text{nozzle}}\) and \(T_{\text{syringe}}\) of 25°C and nozzle diameters of 1 and 1.5 mm.

Circular patterns with a diameter of 6 cm were printed in a single layer to examine the coiling behavior (Figure \ref{fig:methods}a). To study the porosity and fracture behavior, bulk structures consisting of printed straight lines with a length of 10 cm were printed in various numbers of layers, resulting in a height of approximately 1.7 cm. The porosity of said structures was adjusted by manipulating the flow rate at constant nozzle speeds, with increased flow rates leading to a higher porosity (\cite{Willemstein2022}; SI A.2). All of the above designs were created in Gcode Creator, a slicing software developed by TNO, and were printed with varying printer settings to generate different coiling patterns and or porosities. The designs were printed on top of fitted sheets of baking paper that were assumed to be smooth and planar, which were fastened to the printer platform with magnets to prevent slippage. 

Photos and videos of the LRC-3DFP process were captured with a 12MP, 30fps 1080p cellphone camera. In some experiments, the videos captured a lower number of frames per second, which was accounted for in the later calculations. Additional images were captured with a Sony A7 II DSLR camera. For the single-pass designs, ImageJ was used to measure the coil diameters (2R) and track width, with averages of five measurements taken per sample. For designs that were printed in multiple layers, the aforementioned parameters were measured using calipers. The angular coiling frequency (\(\Omega\)) describes the time required for the formation of one coil and was established by analyzing video sequences of the coil formations in Blender (video editor). By counting the number of frames required for the formation of up to five consecutive coils, (\(\Omega\)) of a single loop was established. It is important to note that we assumed the coils to be perfectly circular with a radius of R for our calculations, while they often deviated from the perfect circle and formed slightly teardrop-shaped patterns. A zoomed-in of the printing of the coiled samples and relevant parameters is presented in Figure \ref{fig:methods}a. 

\subsection{Post-processing of printed structures}
Pea and cookie structures printed with a 1.5 mm nozzle were either baked in a convection oven (Leventi Bakermat) at 160°C for 6 minutes, or in an air fryer (Philips) at 160°C for 3 minutes. These baking times were required to produce fully baked structures with moisture contents of 2.91 ± 0.16 and 2.65 ± 0.74 \% wet basis (wb) for cookie and pea ink, respectively. These baking times were halved for the 1.0 mm nozzle. Before baking, the printed structures were stored in closed containers at a refrigerated temperature of 5 $\pm$ 1°C to prevent water evaporation. This storage method was maintained for up to one day to minimize the spreading of the inks during baking.

\subsection{Intrinsic ink properties}
The density (\(\rho\)) of the inks was determined by dividing the mass (\(m\)) of the ink by its known volume (\(V\)), as expressed by \(\rho = \frac{m}{V}\). The water activity (\(A_w\)) of the inks and baked structures was measured with an Aqualab dew point activity meter. In order to establish the material’s moisture content (MC), approximately 1 g of the sample was left in an oven at 104°C for at least 24h for the moisture to evaporate, after which it was cooled down to room temperature in a desiccator for an additional 20 min. MC was expressed in terms of \% on wet base. Both \(A_w\) and MC measurements were performed in duplicate.

The rheological properties of the inks were analyzed with an Anton Paar Rheometer MCR302 fitted with a serrated 25 mm plate-plate configuration and a gap distance of 1 mm. Approximately 2 mL of ink was used per measurement, and all excess material was removed from the geometry prior to analysis. Steady shear measurements were taken, and the shear stress was recorded for shear rates between \(0.0001\) and \(1 \ \text{s}^{-1}\) and back. \(1 \ \text{s}^{-1}\) was chosen as the maximum shear rate because the materials tended to fracture and leave the boundaries of the geometry at higher shear rates. The resulting curves were fitted to the Herschel-Bulkley model, a common method to determine the yield stress \(\sigma_{\text{yield}}\) \cite{Dinkgreve2016}. Elastic (G') and viscous (G'') moduli were determined by amplitude sweeps in oscillatory mode at a frequency of 0.5 Hz, \(0.001\)-\(100\%\) strain, and were also used to establish the yield stress. All measurements were conducted at \(25^\circ \text{C}\), all samples were allowed to rest for 3 minutes before rheological testing, and all analyses were performed in duplicate.

\subsection{Structural porosity}
The external, macrostructural porosity (\(\varnothing_{\text{macro}}\)) of the coiled structures is defined as the air fraction between the coiled filaments. It was calculated as \(\varnothing_{\text{macro}} = \frac{V_{\text{product}} - V_{\text{filament}}}{V_{\text{product}}}\), where \(V_{\text{product}}\) is the volume of the outer dimensions of the entire printed structure, and \(V_{\text{filament}}\) is the volume of the printed filaments. \(V_{\text{product}}\) was determined based on the outer dimensions (length, width, height) of the samples, which were assumed to be perfectly rectangular (Figure \ref{fig:methods}b), while \(V_{\text{filament}}\) was obtained by dividing the weight of the sample by the density of the ink. It was assumed that the porosity remained the same before and after post-processing, as the samples underwent uniform shrinkage. This method for establishing porosity was confirmed by analogous reconstructions of the structures through X-ray computed tomography (see \ref{sec:xrt}).

\subsection{X-ray computed tomography and image analysis}\label{sec:xrt}
By using a GE Phoenix v\textbar tome\textbar x tomographer (General Electric), X-ray computed tomography (XRT) was used to visualize the internal and external structure of the post-processed structures. The complete baked structures and individual filament strands were fastened vertically and a full scan consisting of 1500 projections over 360° was taken. The system contained two X-ray sources and a 180 kV nano focus tube with a tungsten target was employed. X-rays were produced with a voltage of 50 kV and a current of 800 $\mu$A.

The images were recorded by a GE Dynamic41\textbar 200 detector array with 2024 x 2024 pixels (pixel size 200 $\mu$m), that was located 815 mm from the X-ray source. The 3D printed object was placed approximately 130 mm from the X-ray source, resulting in a spatial resolution of resulting in a spatial resolution of $>$2 $\mu$m. The first recorded image was skipped, and the saved projection was the average of 3 images, where every image was obtained over 131ms exposure time. Subsequently, GE reconstruction software version 2.10.1 – RTM was used to calculate and generate the 3D structure via back projection and to compute the volume of the full sample, the material matrix, and the air within the matrix. With this data, the more accurate structural porosity ($\varnothing_{\text{XRTmacro}}$) was established by extraction of the exact outer boundary determined by netting methods and solid matrix volumes from the XRT data, which allowed for a more accurate quantification of the structural porosity. Similarly, by analyzing the air within the matrix, we could compute the microstructural or bulk porosity ($\varnothing_{\text{XRTmicro}}$) of the filaments.

\subsection{Instrumental texture analysis}\label{sec:Textural properties}
Cutting tests on the post-processed structures were carried out with a TA.XT Plus Texture Analyzer (Stable Micro Systems) with a Warner Bratzler blade of 2.9 mm width. A load cell of 30 kg and a trigger force of 0.1 N were used to cut the material at a rate of 5 mm/s up to 95\% strain (\(\varepsilon\)). Texture analysis was performed in triplicate for two independent duplicate samples, leading to 6 datasets per variation. Subsequently, square samples (Figure \ref{fig:methods}b) were subjected to a compression test \cite{bourne1978, kim2012}. These were compressed up to 80\% at a rate of 5 mm/s using a circular probe with a diameter of 4.5 cm. The cutting and compression pressure, \(P\), is given by the force divided over the effective area of the blade (cutting area).

Textural parameters were extracted from the cutting and compression responses, as illustrated in Figure \ref{fig:methods}c. Hardness is defined as the critical pressure of the highest peak (\(P_c\)), brittleness or fracturability is defined as the strain at the highest peak (\(\varepsilon_c\)), and the effective Young’s modulus is used as a measure of initial stiffness, given by the gradient between the pressure and respective strain of the cutting test, following \(E_{\text{eff}} = \delta P / \delta \varepsilon \approx P_c / \varepsilon_c\) \cite{wekwete2008, zoulias2000}. As \(P_c\) could not be obtained from the compression tests, strains in the linear regime of \(\varepsilon < 0.01\) were taken alongside the corresponding \(P\) for the determination of \(E_{\text{eff}}\).

\subsection{Major and meaningful fracture identification}

The mechanical responses in both cutting and compression tests exhibited significant fluctuations, as expected from heterogeneous fragile structures, resulting in numerous maxima and minima in the force-strain curve. We identified two sets of peak values for analysis, (i) major fractures and (ii) meaningful fractures.

\begin{figure}[!h]
	\centering
	\includegraphics[width=0.94\textwidth]{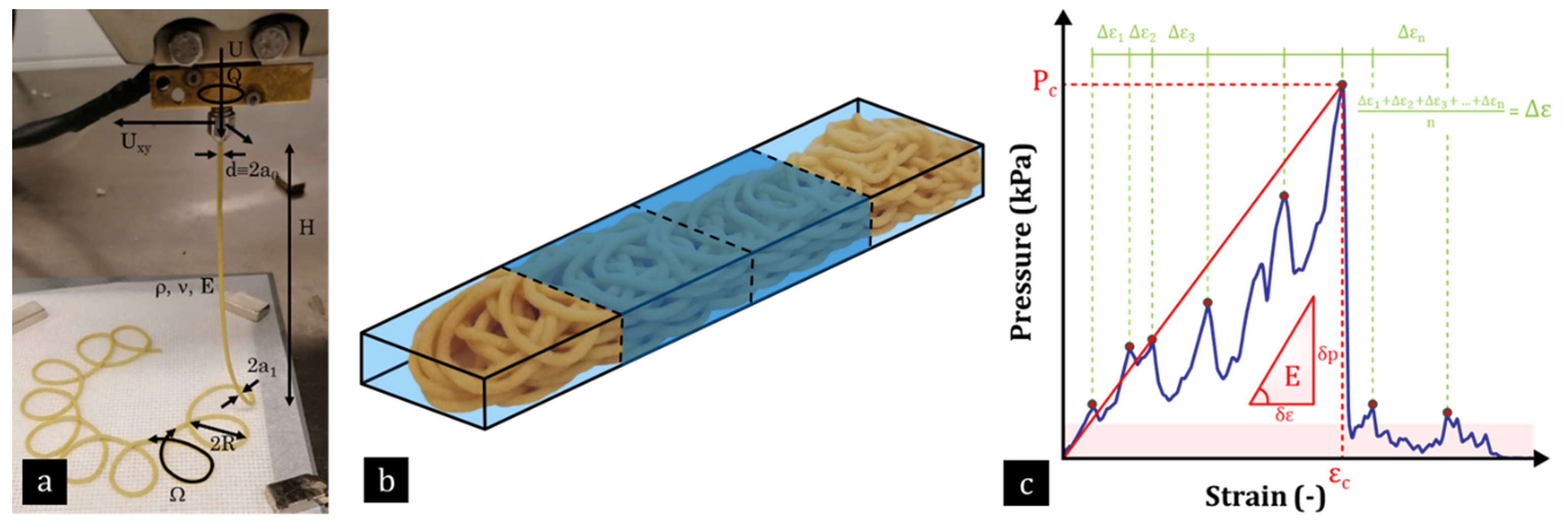}
	\caption{
		Overview and methodologies: 
		(a) Graphic depiction of the printing and material parameters that are important during coiling. 
		(b) Sample volume determined by measurement of the square outer boundaries. The dashed lines represent the cutting cross-sections made with the texture analyzer blade for the cutting tests, and the highlighted cubes that were formed as a result of the cutting were used for the compression test. 
		(c) Example of mechanical responses (here upon cutting) and quantification of the textural properties.}
	\label{fig:methods}
\end{figure}

Major fractures are the peak values, maxima, that surpass \(n\) neighboring data points on each side, identified using the "scipy.signal.argrelextrema" method in Python, with \(n\) consistently set to 5 for all samples. These major peaks represent the fractures with significant contributions to the mechanical response, which are separated by \(\Delta \varepsilon\) strain (red points in Figure \ref{fig:methods}c). Peak forces under 1 N were filtered out in all analyses as they are assumed to play an insignificant role in the textural properties (highlighted red area in Figure \ref{fig:methods}c) \cite{Sowman2010}. 

Meaningful fractures are maxima, which are higher than a threshold (weakest fracture force) from the next minima. To determine the meaningful fractures, we first calculated the difference in the force response between each maximum and the subsequent minima (fracture force, \(\Delta F\)). We then discarded \(\Delta F\) values smaller than the lowest meaningful critical fracture force, \(\Delta F_{\text{min}}\). This value signifies the lowest possible fracture force, setting a threshold for noise in the mechanical responses. Essentially, \(\Delta F_{\text{min}}\) is equivalent to the highest fracture force for cutting a single filament. For the baked cookie filament with a diameter thickness of 1 mm and 1.5 mm, the highest fracture force (\(\Delta F_{\text{min}}\)) was 0.6 N and 1 N, respectively (See SI D.2). As typical force detection threshold for human is about 1 N \cite{Sowman2010}, we estimated \(\Delta F_{\text{min}}\) to be 1 N for all samples in our analysis. We finally calculated the dimensionless meaningful fracture by $\Delta f=\Delta F/\Delta F_{\text{min}}$.

\section{Results and Discussion}

\subsection{Rheology and printability of 3DFP inks}

\begin{figure}[!h]
	\centering
	\includegraphics[width=0.9\textwidth]{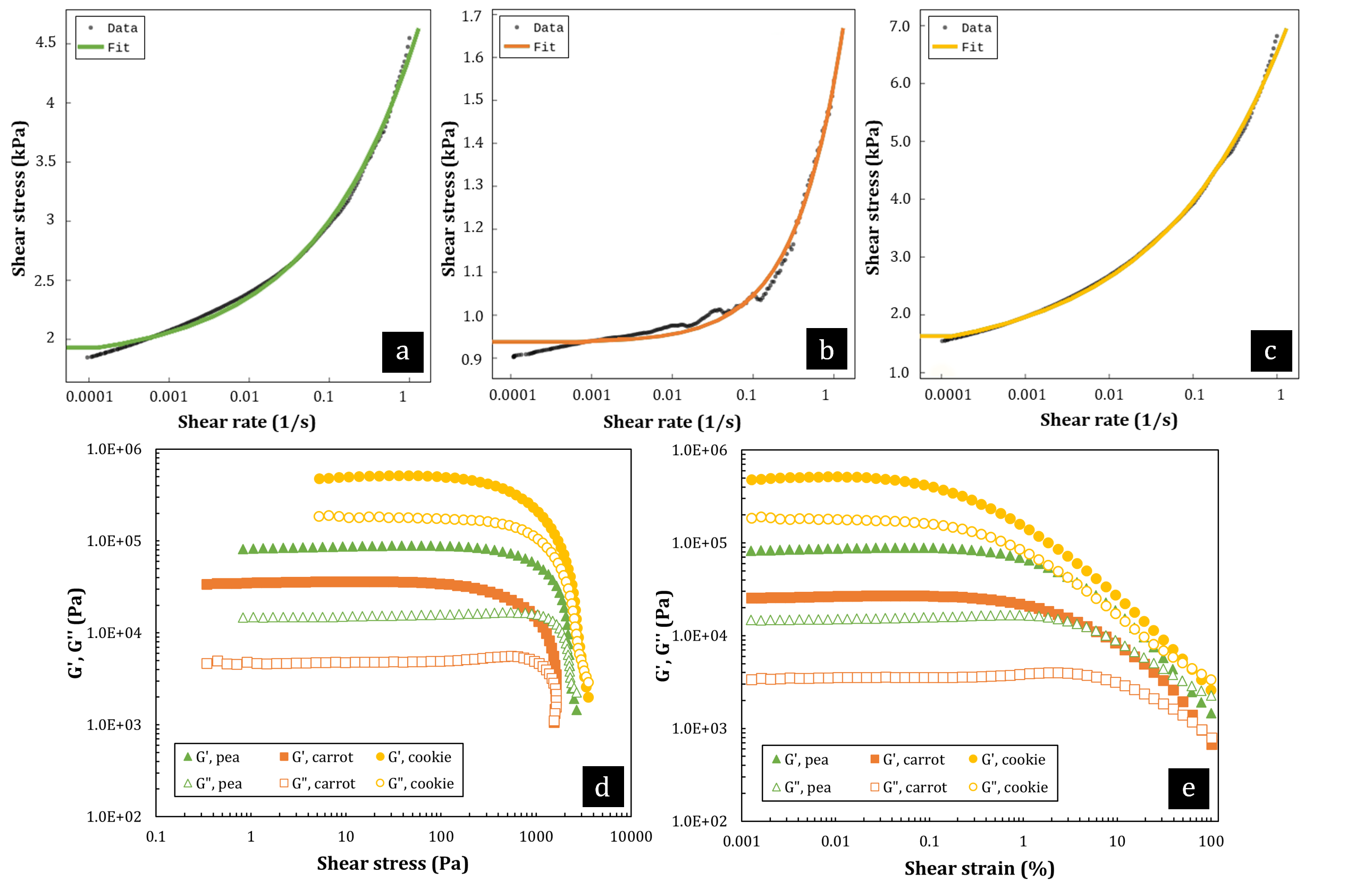}
	\caption{Rheological properties of the 3DFP inks: Shear stress response as a function of the applied shear rate upon steady shear sweeps of 
		(a) pea (green),
		(b) carrot (orange), and
		(c) cookie (yellow) inks fit with the Herschel-Bulkley model. 
		Viscoelastic moduli versus (d) shear stress and (e) shear strain; G'=solid symbols, G"=open symbols.
	}
	\label{fig:rheology}
\end{figure}

First, we characterize the rheological properties of the inks to gain insight into their printability. One ink property that is of main importance during printing is the yield stress ($\sigma_{\text{yield}}$), as the force exerted by the printer must be able to exceed $\sigma_{\text{yield}}$ so it can flow through the nozzle. In Figures \ref{fig:rheology}a-c, the stress-strain curves obtained from steady shear sweeps are fitted with the Herschel-Bulkley model for pea, carrot, and cookie samples, respectively. Pea ink has the highest $\sigma_{\text{yield}}$ (1.85 kPa), closely followed by cookie (1.55 kPa) and carrot (0.91 kPa). The values of $G'$ and $G''$ obtained from amplitude sweep tests are presented as a function of shear stress in Figure \ref{fig:rheology}d and as a function of shear strain in Figure \ref{fig:rheology}e. These results show that the cookie and carrot inks show the highest and lowest $G'$ and $G''$, respectively. We identify similar results for $\sigma_{\text{yield}}$ based on Figure \ref{fig:rheology}d, reported in SI Table B.1. The yield stress of all inks except carrot fall outside the range of 500-1500 Pa reported in literature for printable materials \cite{Liu2019, Outrequin2023}. Yet, these values are highly dependent on the capacity of the printer’s stepper motor. During experiments at high flow rates, the force exerted by the motor on the syringe occasionally went up to 600 N, which may be higher than what most food 3D printers are capable of.

Extended details on rheology of the inks, including a validation of their liquid-like coiling behavior, stability, and large amplitude oscillatory shear (LAOS) responses, are discussed in SI B.

\subsection{Characterization of the coiling behavior of 3DFP inks}

\subsubsection{Effects of printing height on coiling}\label{sec:effects of printing height on coiling}
\begin{figure}[!h]
	\centering
	\includegraphics[width=0.9\textwidth]{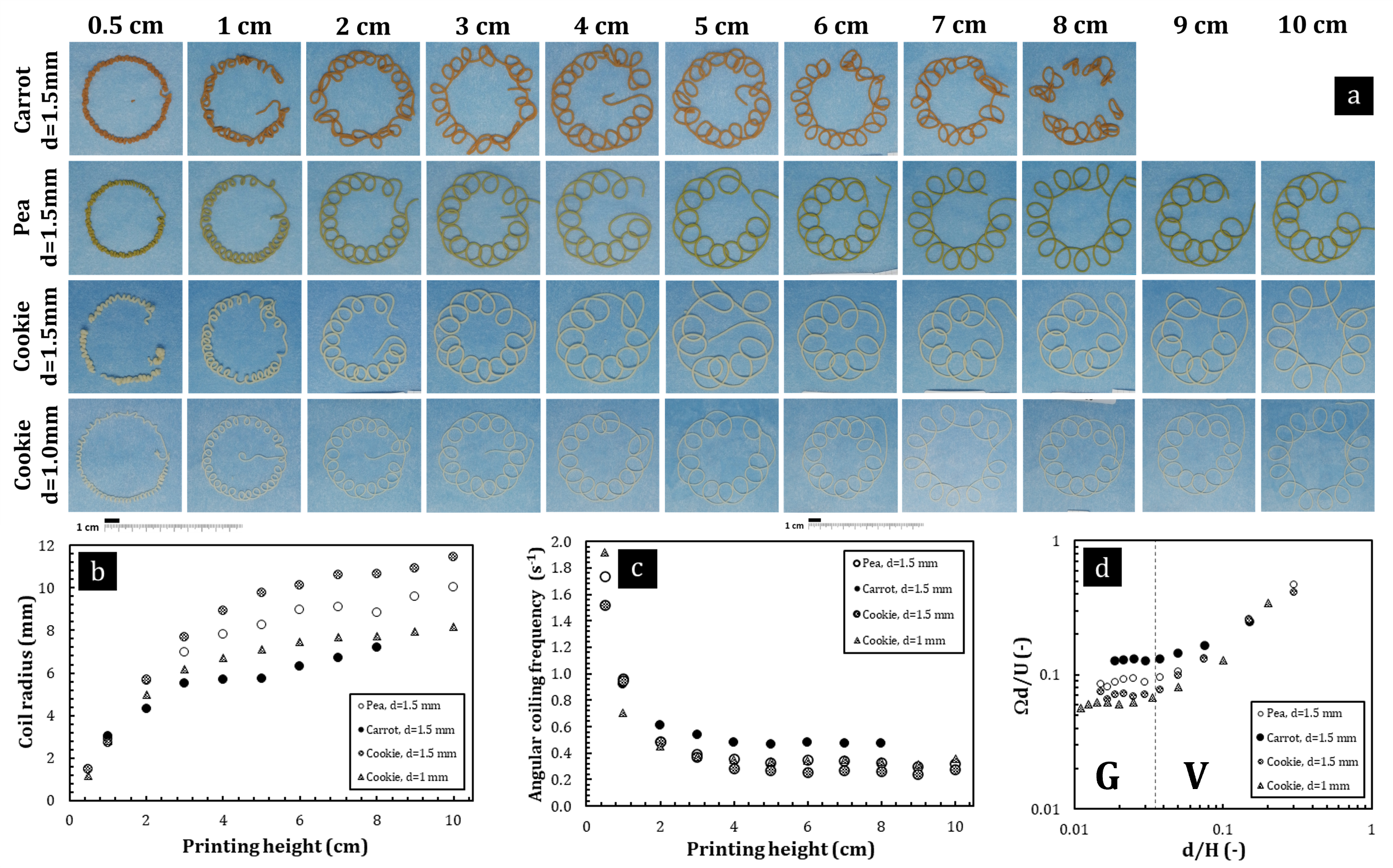}
	\caption{
		The effect of printing height on the coiling behavior: 
		(a) sequence of images showing carrot, pea and cookie inks printed with a 1.5 mm nozzle at a flow rate of 9.79 mm\(^3\)/s, and cookie ink printed with a nozzle diameter of 1.0 mm at 4.35 mm\(^3\)/s; 
		(b, c) the corresponding coil radii and frequencies; 
		(d) the nondimensional coiling frequency \(\Omega d/U\) versus the nondimensional height \(d/H\) in a logarithmic scale. The dashed line at \(d/H=0.03\) represents the boundary between the gravitational (G) and viscous (V) regime.}
	\label{fig:coilingH}
\end{figure}

In Figure \ref{fig:coilingH}a, four series of images display the patterns that emerged from depositing the inks in a circular line with a constant flow rate (9.79 mm\(^3\)/s), with printing heights ranging from 0.5 cm to 10 cm between the nozzle and platform. To characterize the effect of printing height on coiling properties, we quantified the coiling radius (\(R\)) and frequency (\(\Omega\)) for different experiments by analyzing images in Figure \ref{fig:coilingH}a and associated videos; supplementary videos 1 and 2 exemplify the process of coiling of cookie and pea samples.

From Figure \ref{fig:coilingH}b, it can be observed that by increasing the printing height (\(H\)) up to 3 cm, the coils’ radii increase linearly (linear regime) and follow a master curve for all inks and nozzle diameters. This means that \(R\) is controllable in the linear regime, independent of the ink type and nozzle diameter. However, by increasing the printing height further (\(H > 3\) cm), the coil radii diverge from the linear trend and converge to relatively constant radii, with different values depending on the ink type and nozzle size. Therefore, further increasing \(H\) will not increase \(R\) significantly, which makes printing at increased heights not interesting in this study as it does not allow for the creation of larger coils. Figure \ref{fig:coilingH}c depicts the effect of height on the coiling frequency. Initially, the coiling frequency sharply decreases with increasing height (for \(H < 2\) cm), but quickly reaches a minimum value and remains relatively constant thereafter.  

In Figure \ref{fig:coilingH}d, we plot the dimensionless frequency, \(\Omega d/U\), as a function of the inverse of dimensionless height, \(d/H\), as this is a useful representation to identify different coiling regimes \cite{Habibi2007, Maleki2004, Rahmani2011}. At low heights, where \(d/H \ge 0.03\), the data for all samples converge on a line with a slope of 1, representing the viscous regime (V), associated with the linear regime in Figures \ref{fig:coilingH}b and c. In this coiling regime, the coiling frequency is proportional to the inverse of height (\(\Omega \propto H^{-1}\)) and both inertial and gravitational forces are negligible \cite{Habibi2006, Maleki2004}. However, with increasing height, for \(d/H < 0.03\), the dimensionless frequency remains constant (data on a line with a slope of 0), indicating the transition to the gravitational regime (G). The value of the minimum frequency at large heights depends on the ink type and nozzle diameter but falls in the range of \(0.2 < \Omega < 0.6 \text{s}^{-1}\). As shown in Figure \ref{fig:coilingH}a, carrot ink becomes unstable and breaks at \(H \ge 8\) cm. At this critical height, the filament reaches its yield stress, leading to eventual breakage. No critical height was observed for the cookie and pea inks in the heights explored in this study. 

The differences in coiling behavior among individual inks are attributed to the rheological properties of the inks and the nozzle size used. In the gravitational regime in Figure \ref{fig:coilingH}b (\(H > 3\) cm), the coil radius is higher for samples with higher viscosity, which is consistent with previous research on LRC \cite{ribe2012}. In the viscous regime (\(H < 3\) cm), however, the coiling frequency and radius neither depend on the rheological properties nor the nozzle diameter, which has also been confirmed by previous studies \cite{Habibi2007, Maleki2004}. Nonetheless, both parameters are significantly dependent on the mass flow rate upon deposition. Therefore, we will investigate the effect of flow rate in greater detail in the following section.

\subsubsection{Effects of flow rate on coiling}\label{sec:effects of flow rate on coiling}
\begin{figure}[!h]
	\centering
	\includegraphics[width=0.9\textwidth]{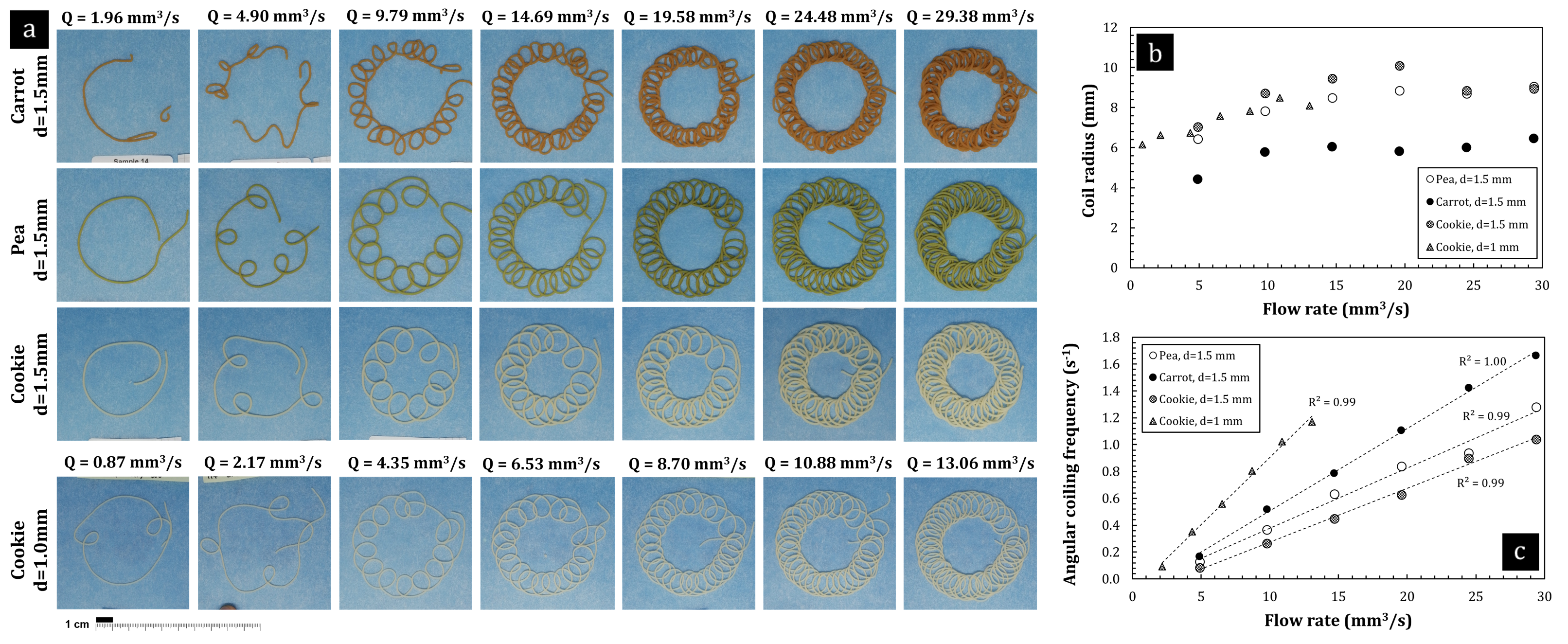}
	\caption{
		The effect of flow rate on coiling behavior: 
		(a) sequence of images showing carrot, pea, and cookie doughs printed with a 1.5 mm nozzle, and cookie dough printed with a nozzle diameter of 1.0 mm, both printed at a height of 4 cm and movement speed of 300 mm/min;
		(b, c) coil radius and frequency as a function of flow rate.}
	\label{fig:coilingQ}
\end{figure}

For the purpose of creating reproducible structures, the translated coiling pattern is preferred, since it is the most stable pattern, allowing for reproducible printing. Figure \ref{fig:coilingQ}a displays the effects of increasing flow rates (\(Q\)) on the coiling behavior while the ink is deposited from a constant height (4 cm). For the first two flow rates studied here (4.9 and 1.96 mm\(^3\)/s), the slow speed of deposited filaments results in unstable and less reproducible coiling patterns \cite{Habibi2011}. Upon increasing the flow rate, coiling starts to occur in alternating or stretched loop patterns. Further increasing of the flow rate results in translated coiling patterns at \(Q \geq 9.79\) mm\(^3\)/s, for all inks printed with a 1.5 mm nozzle, and at \(Q \geq 4.35\) mm\(^3\).

As shown in Figure \ref{fig:coilingQ}b, the coiling radius does not strongly depend on the flow rate, as it remains relatively constant for all ink types. This is in contrast to the findings of \cite{Willemstein2022}, who reported coil radii to decrease as a function of the flow rate, attributed to the shear thinning viscosity of materials upon higher extrusion rates. However, the flow rate range studied here may not have been high enough to investigate similar effects. Figure \ref{fig:coilingQ}c reveals a linear relationship between the coiling frequency and flow rate, which is congruent with literature \cite{mansoori2018, Willemstein2022}. The increasing trend is linear and of a repetitive nature for all samples, but the sample with a lower nozzle diameter (\(d=1\) mm) displays a sharper increase. Increasing the coiling frequency via increasing the flow rate (multiple methods to do so are described in SI C.1) is a valuable method for structural designing, as it corresponds to an increase in the number of coils formed per printed distance, and therefore, also the structural porosity.

\subsubsection{Optimal printing parameters for coiled designs}
Based on the coiling experiments, a printing range was established for the design of structures to fall within the translated coiling regime, ensuring optimal coiling stability and predictability. The ranges presented in Table \ref{tab:printing_parameters} are based on the heights and flow rates tested, so the reported boundary values are approximate. Deposition heights of 4-7 cm yielded stable coils for the carrot ink, and heights of  $>1$ cm
were suitable for the pea and cookie inks, regardless of the nozzle size. Ultimately, for the design of further structures, we used a constant deposition height of 4 cm for all inks to maintain similar coil radii and structure dimensions. For design purposes, we were mainly interested in varying the coil density. This was achieved by varying the flow rate between 3.26 mm\(^3\)/s and 146.89 mm\(^3\)/s as stable coils were formed within these ranges, although exact ranges varied slightly per ink type. The specific upper flow rate boundaries are based on the point after which coiling becomes erratic – and quite literally spins out of control. Moreover, increasing \(Q\) will also slightly affect other sample properties such as the filament diameter due to die swell.

For design purposes, it is also noteworthy to mention that the initiation of coiling (moving bed or nozzle) and the printing path (circular versus square) do not affect \(\Omega\) and \(R\) of the coils (elaborated upon in SI C.2). Thus, different 3D printers can be used to print coils along different paths while maintaining good reproducibility. However, sharp turns in the printing path, especially at low flow rates, can reverse the coiling direction and should be avoided if undesired.

\begin{table}[!htb]
	\centering
	\caption{Optimal printing ranges established for the different inks in order to obtain stable translated coiling patterns. *Coiling was still stable at maximum nozzle height (10 cm) of the printer.}
	\label{tab:printing_parameters}
	\begin{adjustbox}{width=\textwidth, margin=0pt 0pt 0pt 0cm} 
		\begin{tabular}{lll}
			\toprule
			\textbf{Printing parameter} & \textbf{Ink type} & \textbf{Optimal printing range} \\
			\midrule
			\multirow{2}{*}{Height (H)} & Carrot & $4 \, \text{cm} \leq H \leq 7 \, \text{cm}$ \\
			& Pea, cookie ($d = 1.0, 1.5 \, \text{mm}$) & $1 \, \text{cm} \leq H \leq \text{n.d.*}$ \\
			\midrule
			\multirow{4}{*}{Flow rate (Q) at $H = 4 \, \text{cm}$} & Carrot & $9.79 \, \text{mm}^3/\text{s} \leq Q < 97.92 \, \text{mm}^3/\text{s}$ \\
			& Pea & $3.26 \, \text{mm}^3/\text{s} \leq Q < 130.57 \, \text{mm}^3/\text{s}$ \\
			& Cookie ($d = 1.5 \, \text{mm}$) & $3.26 \, \text{mm}^3/\text{s} \leq Q < 146.89 \, \text{mm}^3/\text{s}$ \\
			& Cookie ($d = 1.0 \, \text{mm}$) & $4.35 \, \text{mm}^3/\text{s} \leq Q < 54.40 \, \text{mm}^3/\text{s}$ \\
			\bottomrule
		\end{tabular}
	\end{adjustbox}
\end{table}

\subsection{Using coiling to create porous 3D printed structures}

\subsubsection{Macrostructural porosity}
\begin{figure}[!h]
	\centering
	\includegraphics[width=0.9\textwidth]{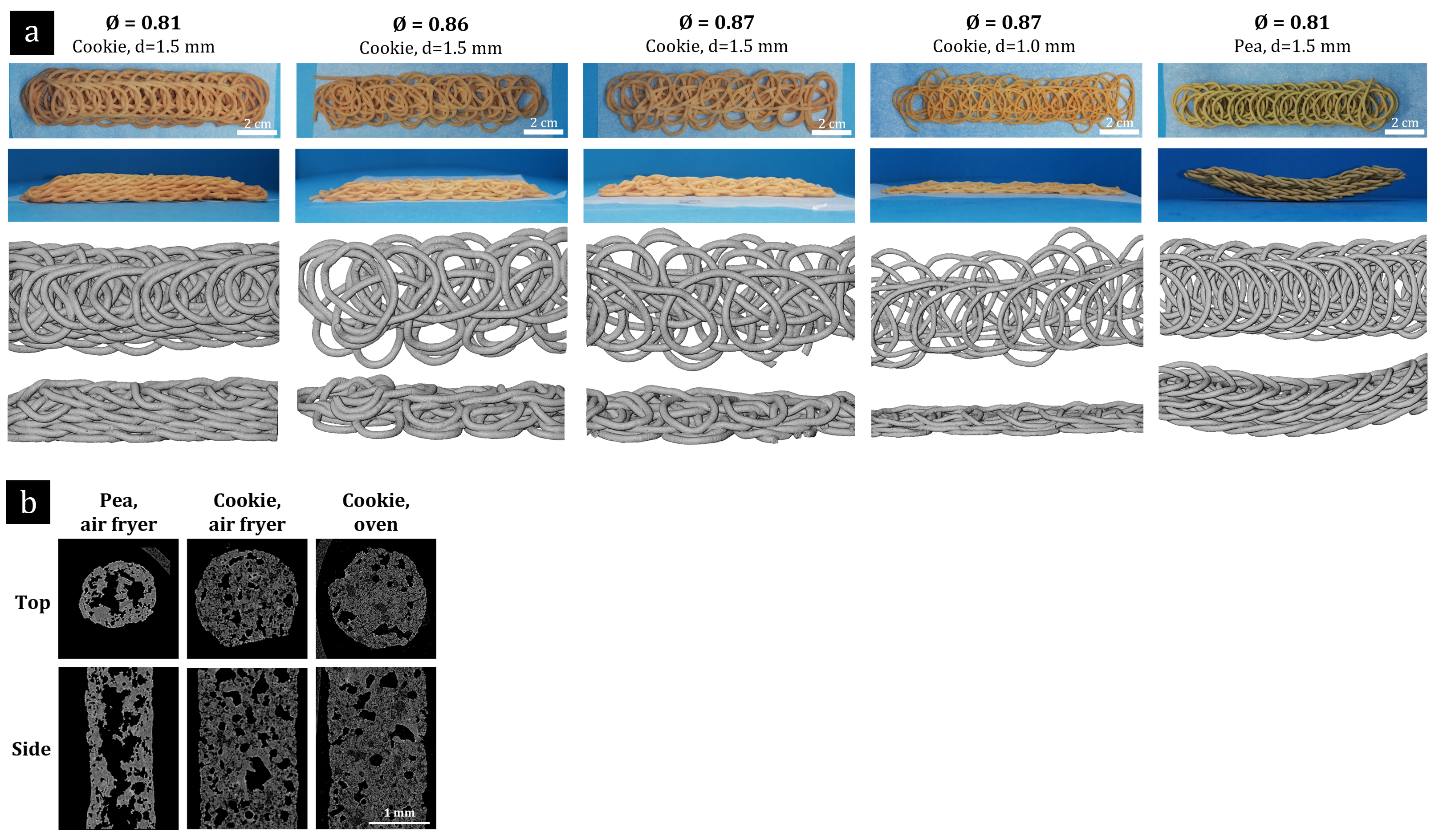}
	\caption{
		Coiled structures: 
		(a) Photographs and reconstructed 3D X-ray tomographs of air fried coiled structures with varying porosity. (b) Top and side-view X-ray tomography scans of post-processed pea and cookie ink strands printed with a 1.5 mm nozzle. The gray areas depict the material matrix, and air is represented by the black areas.}
	\label{fig:xrt}
\end{figure}

With the identified optimal printing ranges for translated coiling, coiled structures with different degrees of overlap between coils, i.e., porosity, were designed from the cookie and pea inks and were subsequently baked in an oven or air fryer. A selection of structures with increasing macrostructural porosities (\(\varnothing_{\text{macro}}\) and \(\varnothing_{\text{XRTmacro}}\)) and their corresponding X-ray tomographs (XRTs), are showcased in Figure \ref{fig:xrt}a. An overview of all coiled structures and the printer settings used to print them is provided in SI C.3. These images effectively capture the distinctive ‘woven’ nature of the coiled structures, although quite some variation exists between the structures. The low porosity samples appear more tightly woven together, with more overlap among strands that stack upon each other rather uniformly, whereas the higher porosity structures consist of more separated loosely overlapped strands. Yet, pea and cookie structures with similar porosities have similar visual appearances with similar coiling patterns and strand overlap, which can be explained by the similar coiling behavior all inks exhibited (sections \ref{sec:effects of printing height on coiling} and \ref{sec:effects of flow rate on coiling}).

Furthermore, consistent differences were found between \(\varnothing_{\text{macro}}\) and \(\varnothing_{\text{XRTmacro}}\), indicating that both approaches accurately represent macrostructural porosity (see SI C.4). However, not all samples were subjected to XRT analyses, so although \(\varnothing_{\text{XRTmacro}}\) is preferred, henceforth only \(\varnothing_{\text{macro}}\) will be presented. It should also be noted that with the printer settings applied, only a relatively high range of porosities (\(0.79 \leq \varnothing_{\text{macro}} \leq 0.90\)) could be attained. This is a range that is very different from that of conventional cookies or biscuits (\(\varnothing_{\text{macro}}\approx0\)) but shows that our LRC-3DFP method allows for the creation of cookie-like products with much higher porosities, potentially affecting the textural and thus sensory properties.

\subsubsection{Microstructural porosity and properties}
In addition to macrostructural porosity (the air fraction between the filaments), the porosity inside the filaments is defined as microstructural or bulk porosity (\(\varnothing_{\text{XRTmicro}}\)). From the XRT images in Figure \ref{fig:xrt}b, it can be observed that large visual differences exist between pea and cookie microstructures, although no large differences were found among \(\varnothing_{\text{XRTmicro}}\) of the samples, regardless of the printer settings or the ink type (see SI C.4). In pea strands, air is predominantly concentrated in a single large pocket in the center of the filament, whereas air pockets in cookie strands are finer and evenly distributed. All strands were printed with a 1.5 mm nozzle, so from the scale, it can be observed that the pea strand shrunk during post-processing, while the size of the cookie strand remained more or less similar. This can be attributed to the high moisture content of the pea dough (53.10±0.03\% wb), resulting in greater water loss during baking, causing it to shrink. The cookie ink, on the other hand, contained less water (15.81±0.15\% wb), and the matrix was able to better hold the small gas bubbles, which is known to be a unique property of wheat-based doughs.

Furthermore, Figure \ref{fig:xrt}b reveals no clear differences among cookie inks baked in an air fryer or oven. To the best of our knowledge, the effects of these post-processing methods on the microstructure of bakery products have not been addressed in literature before. Therefore, these findings qualitatively explain the differences between baking methods, although further analyses targeting air bubble size, shape, and distribution should be completed to validate the observations presented here. Moreover, in this study, we did not intend to change the microstructure of the samples. Yet, it should be noted that the mechanical properties of the food structure also depend on the material properties of the matrix; the ingredient composition and molecular interactions that occur during (post-)processing \cite{Noort2017}.

\subsection{Textural properties of coiled structures with varying porosities}

\begin{figure}[!h]
	\centering
	\includegraphics[width=0.9\textwidth]{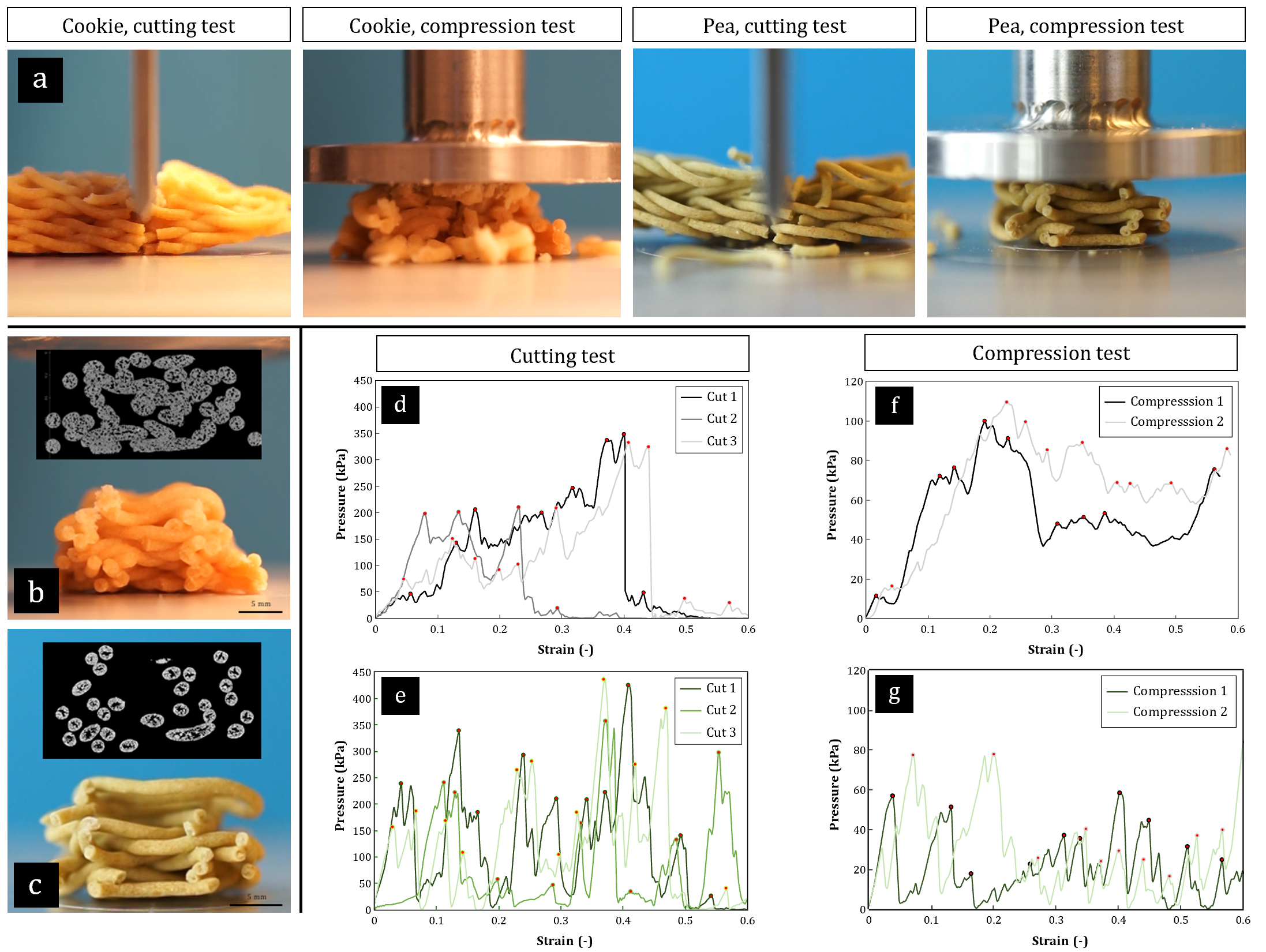}
	\caption{
		Cutting and compression experiments:
		Air fried pea and cookie structures, printed with a nozzle diameter of 1.5 mm, a printing height of 4 cm, and a flow rate of 19.58 mm\(^3\)/s, resulting in a porosity of 0.81. 
		(a) Stills of both structures being subjected to a cutting test (at $\approx$0.5 strain) and a compression test (at ~0.25 strain). 
		(b,c) Photographs and XRT scans of cross-sections of the cookies (top: cookie, bottom: pea). 
		(d,e) Cutting and compression tests for the cookie sample. 
		(f,g) Cutting and compression tests for the pea sample. The red marks depict the mathematically identified main peaks.}
	\label{fig:TA}
\end{figure}

The post-processed structures were subjected to instrumental cutting and compression tests to measure their mechanical properties. In Figure \ref{fig:TA}a, the cutting and compression states of a cookie and pea structure are presented. The cross-section of the cuts and the corresponding XRT images are shown in Figures \ref{fig:TA}b and c for cookie and pea samples, respectively. As the strands of the coiled structures create a network of random-like overlaps, the cross-sections show an irregular pattern of individual filament strands, which is also clearly visible in the XRT images (Figure \ref{fig:xrt}).

We will now further discuss the mechanical response of the systems to cutting and compression. To obtain comparable variables among the differently sized samples, the applied force was scaled by the effective area of the sample (including the void) that was in contact with the probe, resulting in pressure (kPa), which is represented as a function of the applied strain in Figures \ref{fig:TA}d-g. The irregular distribution of strands becomes evident from the response of the systems showing a series of individual fracture events that seem to follow a degree of randomness for all samples, indicating that many small fractures occur rather than one large one.

In cuts 1 and 3 in Figure \ref{fig:TA}d, an increase in pressure during cutting can be observed until an onset of complete fracture is reached at a strain of around 0.4. Figure \ref{fig:TA}e shows that, upon cutting, structures printed with pea ink present more distinct and jagged peaks and lack a clear point of complete fracture. Upon compression, the cookie samples (Figure \ref{fig:TA}f) exhibit a continuously increasing fracture pressure regime in low compression (strain of $<0.2$), where the pressure does not reduce significantly. However, upon large compression, the pressure falls to almost half of its maximum and fluctuates around that value until the saturation regime is reached for compression strains above $\approx 0.5$. This is not the case for compression of the pea samples presented in Figure \ref{fig:TA}g, which present more jagged peaks that fall back to a value of around 0 kPa, indicating the sequential collapsing of individual layers of strands, which can also be observed in Figure \ref{fig:TA}a.

The differences between the fracture behavior of the cookie and pea systems could be attributed to the fusion of individual filament strands and the differences in the distribution of voids in the filament cross-section. The cookie ink contains fat, causing the individual strands to become more fused during post-processing (Figure \ref{fig:TA}b), while the individual pea filaments remain separated (Figure \ref{fig:TA}c) and have thicker cell walls (Figure \ref{fig:xrt}b), leading to more jagged fracture profiles. Furthermore, this irregular fracture behavior of both systems is unlike conventional 3D printed structures or common food samples, where the bulk composition determines the overall fracture behavior of the structure, which is usually prone to complete fracture upon abrupt buckling \cite{Noort2017}. This type of singular fracture for solid cookie structures is confirmed in SI D.1.

Even though the pressure-strain curves for all samples display an erratic response and stochastic behavior (with several repetitions indicated by the color scales), we demonstrate that the general characteristics of the fracture events are consistent and primarily depend on their porosity. To further analyze the textural properties of the systems, we quantified their hardness, brittleness, and initial stiffness, as defined in \ref{sec:Textural properties}, and compared these parameters between different samples. Besides these commonly used parameters in textural analysis, we also quantified the distance between major fractures and the distribution of the meaningful fractures. These analyses aimed to provide a better understanding of the disordered mechanical response and an insight into the sensory perception of coiled structures.

\subsubsection{Hardness}
\begin{figure}[!h]
	\centering
	\includegraphics[width=0.9\textwidth]{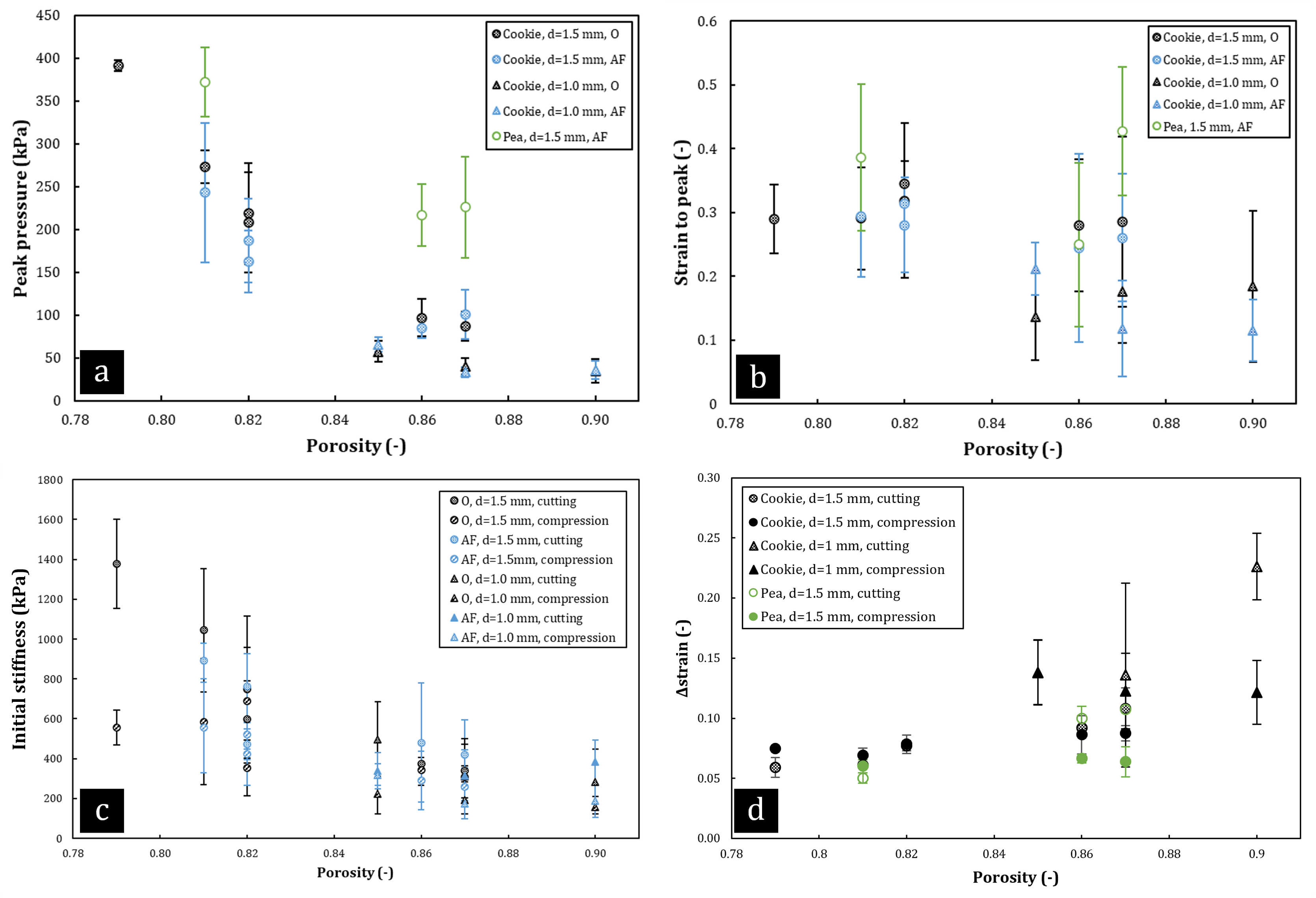}
	\caption{
		Fracture properties:
		(a) Peak pressure determined from cutting tests as a function of porosity, indicative of hardness. 
		(b) Strain at peak force determined from cutting tests as a function of porosity, giving an indication of brittleness. 
		(c) Initial stiffness of the cookie samples determined from cutting and compression tests, as a function of porosity. 
		(d) \textnormal{$\Delta$strain} determined from cutting and compression tests as a function of porosity. All samples were printed with a nozzle with a diameter (d) of 1.0 or 1.5 mm, and were either oven baked (O) or air fried (AF).}
	\label{fig:TA attributes}
\end{figure}

As depicted in Figure \ref{fig:TA attributes}a, the hardness of all 3D printed structures decreases with increasing structural porosity, as a lower peak pressure represents a lower resistance to fracture, resulting in a less hard texture. Reducing hardness as a function of porosity is more pronounced for porosities below 0.84 (\(\varnothing_{\text{macro}}<0.84\)). However, for \(\varnothing_{\text{macro}}>0.84\), the impact of porosity is not significant, and the hardness remains relatively constant. This can be interpreted as the existence of two regimes of mechanical responses: i) the low-porous regime, where hardness decreases with increasing porosity; ii) the high-porous regime, where hardness remains constant as a function of porosity and only the microstructure affects the hardness. The existence of these two dominant regimes is consistent across all samples. Besides porosity, the composition of the sample plays a significant role in determining the hardness of the system, as the pea samples show the hardest overall texture. Furthermore, no significant differences can be observed in terms of hardness between the oven-baked or air-fried cookie samples.

The hardness of all coiled samples is comparable to those previously published for regular cookies, albeit on the low side, thereby confirming their validity \cite{baltsavias1999, kim2012, saleem2005}. In these studies, differences in hardness among samples are mainly attributed to differences in material composition, represented by moisture content (MC) and water activity (\(A_w\)). However, the MC and \(A_w\) of the baked cookie and pea structures are relatively similar (MC: \(2.65\pm0.91\), \(2.91\pm0.17\%\) wb; \(A_w: 0.25\pm0.10\), \(0.18\pm0.05\), for pea and cookie, respectively), leading to the postulation that in this study, hardness is dependent on the coiled structure’s geometry and the post-processed ink’s microstructure as evaluated by the XRT scans. Based on the structural geometry, there are fewer points of overlap between strands of structures with higher porosities, leading to a lower resistance to fracture and thus, a lower hardness. This same description holds true for the samples printed with a smaller diameter nozzle, which also leads to a lower resistance to fracture.

\subsubsection{Brittleness}
The strain up to the main point of fracture before the major collapse of a material indicates its brittleness, i.e., the ease of fracture. In more brittle materials, the fracture will form more quickly, which means the measured strain until the major collapse of the material will be lower. From Figure \ref{fig:TA attributes}b, it can be observed that the degree of brittleness is relatively similar across different structural porosities, with pea samples being the least brittle and the cookie samples (printed with a 1.0 mm nozzle) being the most brittle. Therefore, differences in the brittleness of different inks could be attributable to the differences in their microstructure, such as the distribution of air within the filament cross-section.

\subsubsection{Initial stiffness}
The initial stiffness of the cookie samples, represented by the effective Young’s modulus, follows a trend similar to that of the hardness and decreases with increasing structural porosity and decreasing nozzle sizes, as shown in Figure \ref{fig:TA attributes}c. The initial stiffness obtained from the cutting tests nonlinearly decreases as a function of porosity. However, the initial stiffness based on the compression test remains relatively constant over the entire porosity range. The initial stiffness determined by the compression tests is consistently lower than those obtained from cutting tests due to the different nature of fractures created through cutting and compression. These differences offer insights into anticipating differences in sensory perception of the samples upon biting with frontal teeth or chewing with back teeth, which are expected to display similarities to cutting and compressing, respectively \cite{Varela2009}. For the pea samples, no clear trend could be observed for the initial stiffness, although overall, it was higher than that of the cookies. Due to the high standard deviations and resulting irreproducibility, this data was not included in this study, but the graph is presented in SI D.3. Additionally, no significant differences in terms of hardness, brittleness, or initial stiffness were found between nearly all cookie structures baked in an oven and air fryer. This lack of difference in post-processing method can be attributed to the similar macro- and microstructural attributes of the systems.

\subsubsection{Strain interval between major fracture events}
For further analysis, the average distance (i.e., strain) between the major peaks, henceforth referred to as \(\Delta\text{strain}\), was determined. Similarly to brittleness and stiffness trends, we can distinguish a low-porous and high-porous regime for porosities lower and higher than \(\varnothing_{\text{macro}} = 0.84\) (Figure \ref{fig:TA attributes}d). In the low-porous regime, all samples, regardless of the test type (compression or cutting), their composition, and nozzle diameter, display a constant \(\Delta\text{strain} \approx 0.06\). This observation highlights that one fracture event will occur in every 6\% compression strain, which can be directly related to the layers formed during 3D printing. However, in the high-porous regime, \(\Delta\text{strain}\) does not exhibit a clear trend with respect to porosity and significantly varies depending on the sample and test type. This reveals the dominance of unexpected fracture events upon compression in the high-porous regime and can be attributed to the different type of fracture that occurs during compression, where multiple layers get crushed simultaneously or one by one. Similar results for pea and cookie systems indicate that \(\Delta\text{strain}\) is mainly dependent on the geometry of the coiled structure, and not on the composition. Overall, these observations suggest practical strategies for designing customized sensory perceptions by designing porosity in coiled systems through an interplay between flow rate and printing speed.

\subsubsection{Number of fracture events}
\begin{figure}[!h]
	\centering
	\includegraphics[width=0.4\textwidth]{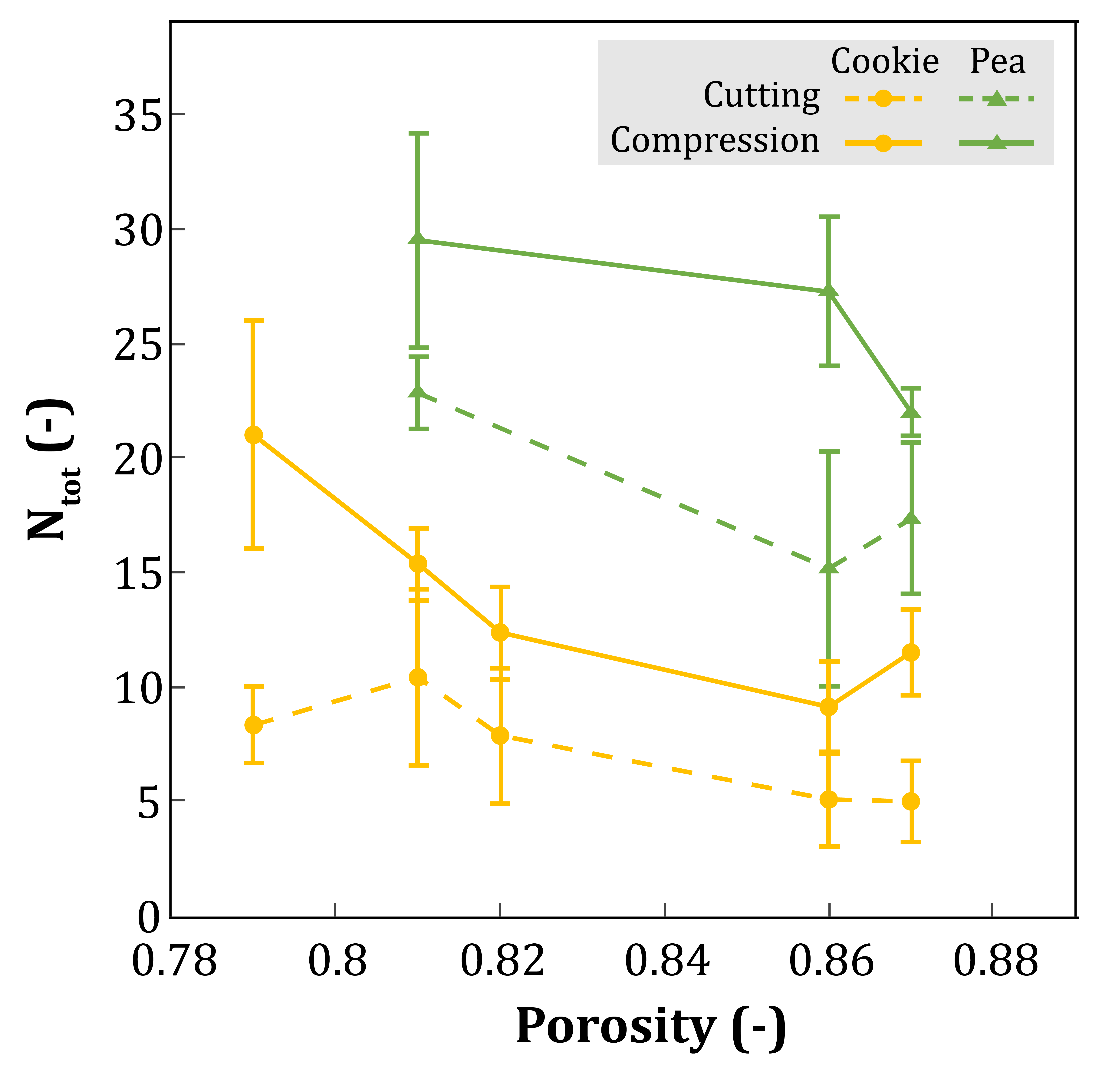}
	\caption{
		The total number of fractures as a function of porosity for cookie (yellow) and pea (green) samples upon compression (solid) and cutting (dash) procedures applied at different sections of the same sample.}
	\label{fig:crunchiness}
\end{figure}

Crunchiness is usually linked to the number of sensible fracture events, which are characterized by the jaggedness of the fracture profile \cite{Suwonsichon1998, Corradini2006}. In order to understand the crunchiness of the coiled systems upon compression and cutting processes, we investigate the total number of meaningful fractures, \(N_{\text{tot}}\), for different samples printed with a 1.5 mm nozzle upon cutting and compression. In Figure \ref{fig:crunchiness}, the total number of fractures as a function of porosity is illustrated. This graph reveals that the pea samples are crunchier than the cookie samples. Moreover, compression results in a higher number of fracture events than cutting, which is expected since during the cutting process the stress is localized on a cutting line while during the compression test, it is distributed over a larger region. Overall, crunchiness tends to decline with increasing porosity, but considering the deviations in the data that are the result of the random nature of the fractures, this is not generically applicable. Despite the random-like behavior of the system, investigating the fracture force distribution reveals additional details and underlying generic mechanisms.

\subsubsection{Fracture magnitude distribution}
\begin{figure}[!h]
	\centering
	\includegraphics[width=0.9\textwidth]{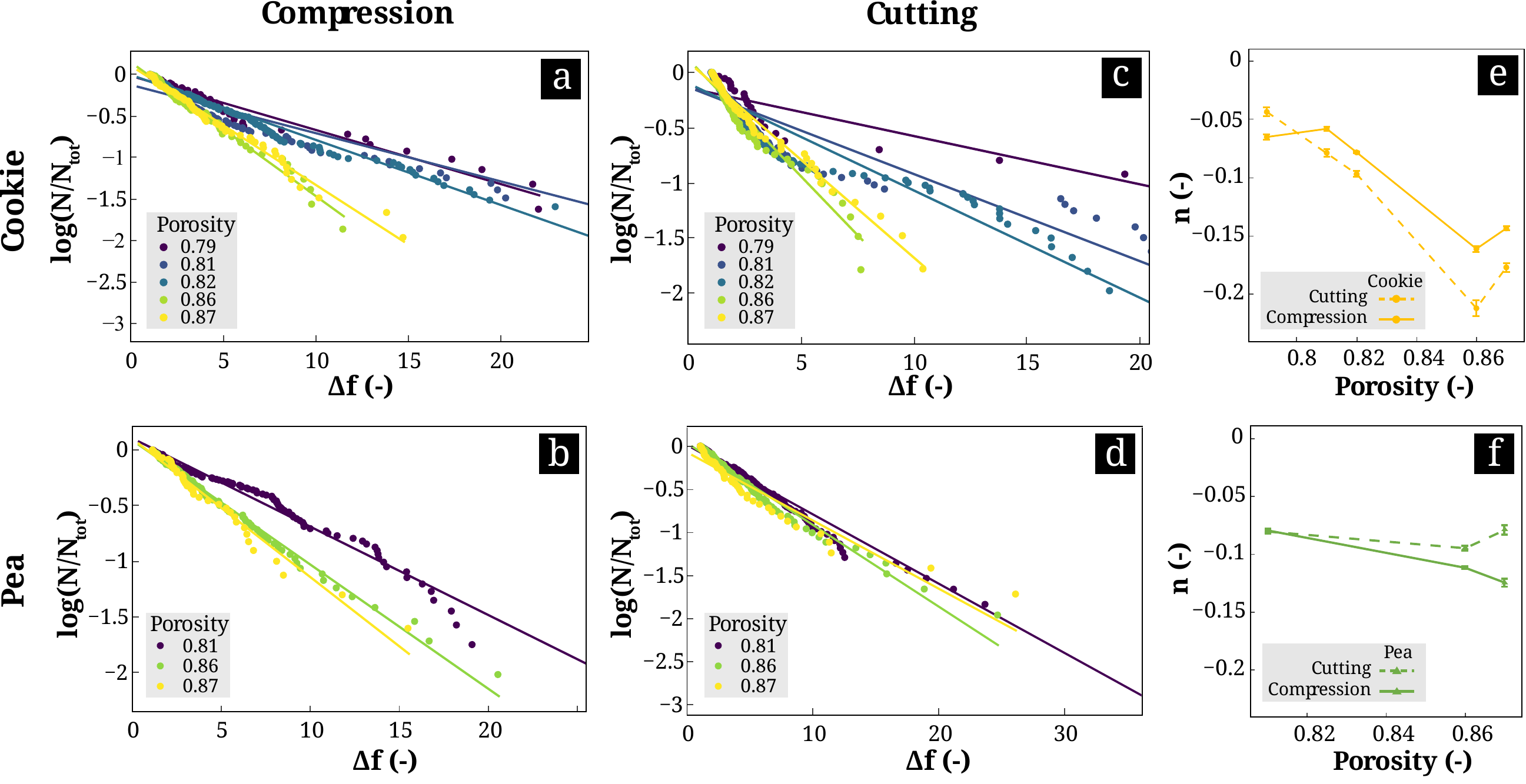}
	\caption{
		Distribution of the fracture magnitude: 
		(a--d) The logarithm of the cumulative occurrence equivalent to or greater than each fracture magnitude, rescaled by the total number of fractures, is illustrated as a function of the dimensionless fracture magnitude for combined oven-baked and air-fried samples during compression (a, cookie; b, pea) and cutting (c, cookie; d, pea) experiments.
		(e, f) The power component in the power-low distribution (slope in log-linear plots) for cookie (e) and pea (f) samples fractured via compression (solid) and cutting (dash) experiments; error bars are based on the deviation of the data from the fitting line.}
	\label{fig:frequency distribution}
\end{figure}

To understand the fracture behavior of coiled structures within a broader context, we analyzed the distribution of the rescaled meaningful fracture force, $\Delta f$. In Figures \ref{fig:frequency distribution}a-d, we illustrate the rescaled number of fracture events equal and larger than each fracture magnitude as a function of the magnitude in a linear-log scale, upon compression (cookie, Figure \ref{fig:frequency distribution}a; and pea, Figure \ref{fig:frequency distribution}b) and cutting (cookie, Figure \ref{fig:frequency distribution}c; and pea, Figure \ref{fig:frequency distribution}d) samples. A generic power-law scaling relationship in the distribution of the fracture force magnitude is given by $\log ({N}/{N_\text{tot}}) = a + n \Delta f$, which is known as the Gutenberg–Richter law \cite{Gutenberg1956}. This law captures the typical characteristic of fracture, and while initially introduced in seismology, it has also been identified in various statistical systems in nature, such as acoustic emission \cite{Carpinteri2006, Golitsyn2013}. Moreover, to compare the fracture force distribution across the samples, the slopes of the fitting lines in Figures \ref{fig:frequency distribution}a-d, which represent $n$, are presented as a function of porosity in Figures \ref{fig:frequency distribution}e (cookie) and f (pea).

Comparing the occurrence of fracture magnitudes between cookie and pea samples reveals distinct characteristics that possibly stem from the coiled system's composition. The cookie samples exhibit a more scattered behavior and a stronger dependence on porosity, particularly evident in the cutting experiment (Figure \ref{fig:frequency distribution}b). For the cookie samples, the value of $n$ declines as a function of porosity in both compression and cutting tests. However, the pea samples demonstrate a more consistent scaling behavior in both compression and cutting tests (Figures \ref{fig:frequency distribution}b, d), exhibiting $n \approx -0.1$ for all samples. These differences may be attributed to the difference in the brittleness of their composition. The difference in fracture profile can be qualitatively observed in the supplementary videos 3 and 4, respectively, upon compression and cutting tests. Our present analysis offers a statistical approach for analyzing fractures in edible systems, which could provide insight into understanding relevant sensory perception.

\subsubsection{Sensory anticipation}
The key question remaining is how individuals would perceive the sensory response of the coiled structures and would distinguish between various fracture profiles. Addressing this question remains beyond the scope of this research; however, we will briefly discuss it based on the available literature studies. Previous research found a general human force perception threshold of 1-6 N, influenced by the direction and speed of mastication and the magnitude of the forces \cite{Sowman2010}. Peaks below 1 N were filtered out in the current study, so in theory, all reported forces should be perceivable. However, the sensory perception also depends on the product type, where a higher sensitivity has been reported for harder products with significant force differences between fracture peaks, but also depends on the time, kinetics, and sound emissions of the fractures \cite{Mioche1994, Williams1984, Luyten2004, Primo-martin}. The ranges of fracture force we identified offer some insights into this perception but do not address differences between consecutive peaks. Hence, further research should quantify these peak force differences to understand the relationship between a structure’s coiled geometry and its fracture properties. Following such an approach in combination with sensory panel evaluations in future studies will provide a comprehensive understanding of the corresponding sensory properties of coiled structures, including bolus formation and additional sensory cues.

\section{Conclusion}

To further expand the capabilities of 3D food printing, we presented a new approach that exploits the liquid rope coiling effect by depositing the filament from a high distance. This method enables the creation of coiled food structures with unique morphological features and tunable textural properties, both primarily driven by the porosity of the coiled structure. The coiled structure was controlled by first identifying optimal printing conditions to induce a specific coiling regime, in this case, translated coiling. The coiling regime could be controlled through the 3D printing parameters, including deposition height, printing speed, and volumetric flow rate. Consequently, this allowed us to control the degree of overlap between coiled filaments and hence the porosity of the food structure. By characterizing the mechanical properties and morphology of the coiled food structures, we demonstrated that the mechanical behavior of coiled structures can be engineered by manipulating their porosity based on the printing parameters.  

In addition, we introduced methods of analyzing fracture data and tried to connect these to the structural properties of the samples. We observed that in a range of porosities, increasing the porosity reduced the hardness and initial stiffness of samples. However, porosity had no major effect on the brittleness, implying that this parameter is mostly dependent on the compositional and microstructural differences among the samples. We observed that the number of fracture events, representing the crunchiness, depends on the composition and consistently is higher for compression tests than for cutting tests. Furthermore, fracture force distributions demonstrated a universality following the Gutenberg–Richter scaling law. 

Based on these findings, we have demonstrated that the utilization of liquid rope coiling in 3DFP offers a relatively faster way of food printing and enables tuning textural properties, which reveals the potential of using coiling for further applications of 3DFP in food design and development.

\section*{Acknowledgements} 
We would like to express our gratitude to Mokhtar Adda-Bedia for the insightful discussions. We also thank Gerard Gim\`enez-Ribes for his support in analyzing and interpreting the rheology data.

\section*{Author Contributions Statement}

A.G. and S.J.G. Contributed equally to this work. A.G. M.N. and M.H. supervised and designed the research. S.J.G. conducted the experiments and wrote the original draft. S.A.G. helped with and supervised the 3D printing-related procedures. A.G. and S.J.G performed the data analysis, visualizations, and revising the manuscript. A.G. developed data analysis codes. M.H. acquired funding for the project. M.H., M.N., and S.A.G. contributed to the validation. All authors discussed the results and contributed to establishing the methodology and experimental procedures. All authors have reviewed the results and approved the final version of the manuscript.

\clearpage
\bibliographystyle{elsarticle-num} 
\bibliography{References}

\clearpage

\setcounter{figure}{0}
 	\renewcommand{\figurename}{Supplementary Figure}

\setcounter{section}{0}
    \renewcommand{\thesection}{SI}

\setcounter{table}{0}
    \renewcommand{\tablename}{Supplementary Table}

\section*{Supplementary Information (SI)}
\section{Detailed information on Materials and Methods}

\subsection{Materials} \label{sec:SI Materials}

Toasted green pea flour was made by toasting (120°C, with 100\% RH super-heated steam for 30 minutes) wrinkled green peas (\textit{Pisum sativum L.}) in a Rational convection oven (SSC WE 61). Subsequently, the toasted peas were dehulled and ground in a ZPS 100 mill (Hosokawa-Alpine) at a milling speed of 20000 rpm, zifter speed of 4000 rpm, and air throughput of 50 m\textsuperscript{3}/h, which obtained a flour with a D50 of 24 $\mu$m and D90 of 80 $\mu$m. Carrot powder consisted of dried and milled carrot pulp, as a residue after carrot juice extraction. This food ingredient was kindly supplied by Van Rijsingen Green, Helmond (The Netherlands).

For the preparation of 500 g of cookie dough ink, 130.04 g of shortening (trio puur zacht) was mixed in a Hobart N50 mixer with a stainless-steel B blat beater for two 30s intervals while cleaning the bowl after each interval. Then, 61.02 g of sugar, 1.94 g salt, and 9.53 g of egg white powder were added to the shortening mix, and were mixed at speed 1 for 30s, followed by another 30s at speed 2. After each step, the sides of the bowl were scraped clean again. Next, 41.50 g water was slowly added (about 60s) while mixing at speed 1, after which the mixture was mixed twice for 30s at speed 2, cleaning the bowl after each step. To this mixture, 152.59 g ibis flour, 51.69 g novation 4300 starch, and 51.69 g wheat starch were added, after which it was mixed at speed 1 until the sides of the bowl remained clean (30-60s). Finally, the dough was mixed for 10 seconds at speed 2. Before use, the dough was put in an airtight bag for at least 1 hour to rest in the fridge to ensure proper hydration. Before use, the dough was brought back to room temperature.  

\subsection{Printing parameters} \label{sec:SI printing parameters}

The dimensionless EM describes the ratio between the plunger speed and the printing speed, to correct for the amount of material extrusion. A higher EM leads to an increased material deposition, potentially altering the overall structure and resolution of the print. The EM is, therefore, related to the piston or injection speed (\(U\)), which can be computed as follows:

\[
U = \text{EM} \cdot U_{xy} \cdot \frac{(w_{\text{track}} - h_{\text{layer}}) \cdot h_{\text{layer}} + \pi \cdot \left(\frac{h_{\text{layer}}}{2}\right)^2}{A_{\text{syringe}}}
\]

In this formula, \(h_{\text{layer}}\) is the layer height equal to \(d/2\), \(w_{\text{track}}\) is the track width equal to \(3 \cdot h_{\text{layer}}\), and \(A_{\text{syringe}}\) is the circular area of the syringe body. The volumetric flow rate (\(Q\)) describes the volume of material deposited from the nozzle per unit time and is dependent on the speed of injection and the area from which it is ejected and can be expressed as:

\[
Q = \pi \left(\frac{d}{2}\right)^2 U
\]

\begin{figure}[!h]
    \centering
    \includegraphics[width=0.5\textwidth]{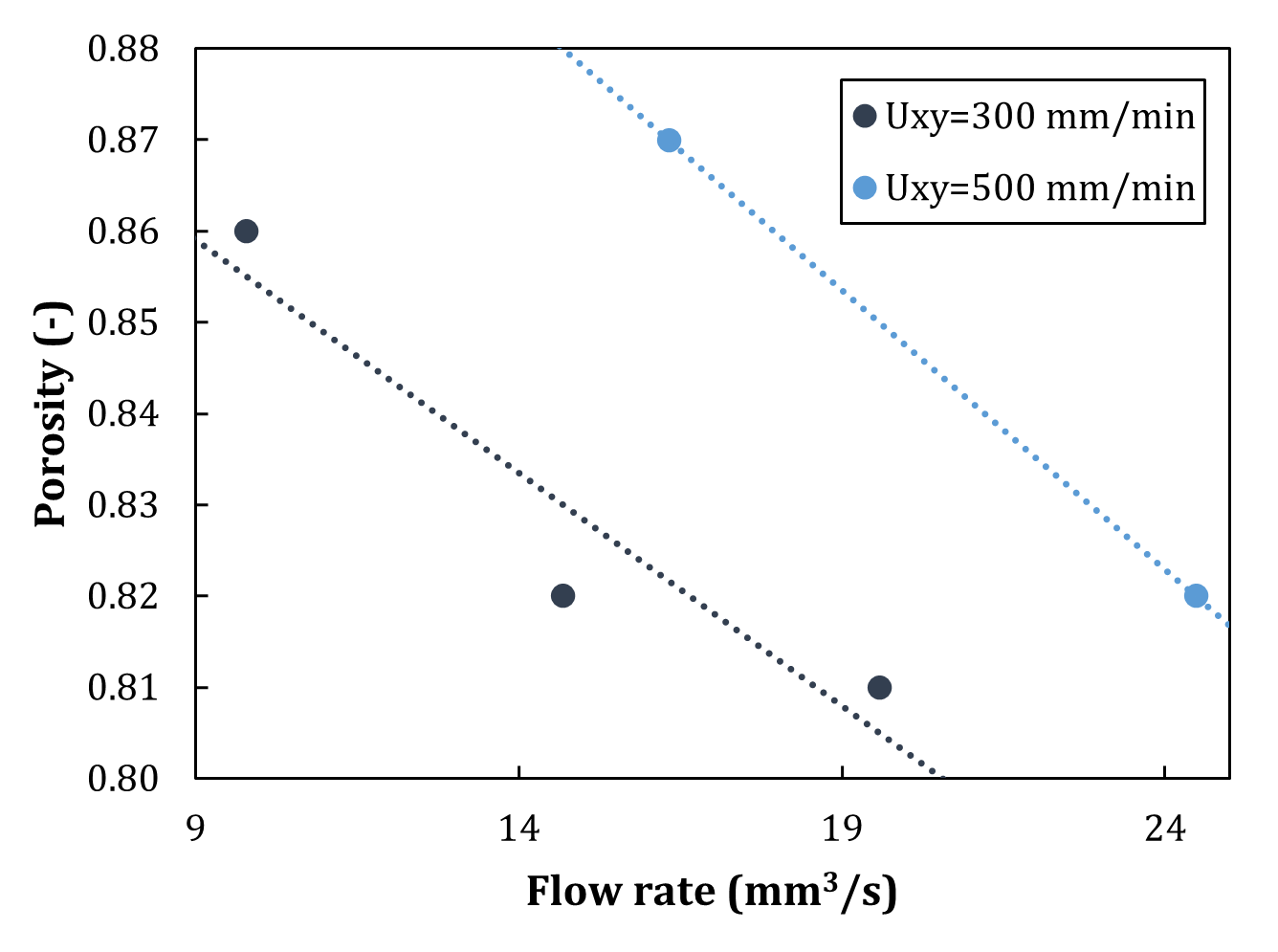}
    \caption{{Porosity versus flow rate at two different nozzle speeds (\(U_{\text{xy}}\))}
    }
    \phantomsection
    \label{fig:SI Q and porosity}
\end{figure}

Furthermore, the flow rate is also directly correlated to the macrostructural porosity of the coiled sample, as can be observed in \ref{fig:SI Q and porosity}. Here it can be observed that higher flow of the ink from the nozzle will result in 3DFP structures with lower porosities, and vice versa for low flow rates.

\section{Further information on the rheological properties of the 3DFP inks}

\subsection{Ink printability and flow behavior}

Besides \(\sigma_{\text{yield}}\) (Supplementary Table \ref{tab: SI Yield stress}), the flow behavior index (\(n\)) and consistency index (\(K\)) can also be retrieved from the Herschel-Bulkley model, and are provided in Supplementary Table \ref{tab: SI rheo ink properties}, along with \(\sigma_{\text{yield}}\). Although these indices are not commonly reported in literature, low \(n\) and \(K\) values are associated with shear-thinning properties, whereas higher values indicate that materials cannot be extruded easily \cite{Outrequin2023, Tejada-Ortigoza2022}. For all inks, \(n < 1\) holds true, signifying a shear-thinning behavior and good extrudability.

\begin{table}[htbp]
\centering
\caption{Yield stress (kPa) of the inks determined by various methods. All values are reported in kPa.}
\phantomsection
\label{tab: SI Yield stress}
\begin{adjustbox}{width=\textwidth} 
\begin{tabular}{@{}lccc@{}}
\toprule 
Ink type & Steady shear & Intersect G’ and G” & \makecell{Extended intersect \\ linear and nonlinear G’} \\
\midrule
Pea & 1.85 & 2.26 & 1.70 \\
Carrot & 0.91 & 0.94 & 1.10 \\
Cookie & 1.55 & 3.09 & 1.60 \\
\bottomrule
\end{tabular}
\end{adjustbox}
\end{table}

\begin{table}[h]
\centering
\caption{Material properties of the inks while being printed with different nozzle sizes.}
\phantomsection
\label{tab: SI rheo ink properties}
\begin{adjustbox}{width=\textwidth} 
\begin{tabular}{@{}lccccccc@{}}
\toprule
Ink type & \( \sigma_{\text{yield}} \) (kPa) & \( n \) (-) & \( K \) (Pa·s\(^n\)) & \( B_{\text{eff}} \) (kPa) & \( \nu_{\text{eff}} \) (\( \times 10^5 \) St) & \( \eta_{\text{eff}} \) (\( \times 10^4 \) Pa·s) \\
\midrule
Pea (d=1.5mm) & 1.85 & 0.33 & 2.50 & 22.84 & 3.09 & 3.37 \\
Carrot (d=1.5mm) & 0.91 & 0.74 & 0.60 & 6.62 & 0.70 & 0.67 \\
Cookie (d=1.5mm) & 1.55 & 0.29 & 5.31 & 46.84 & 7.41 & 8.60 \\
Cookie (d=1.0mm) & 1.55 & 0.29 & 5.31 & 105.40 & 6.01 & 6.97 \\
\bottomrule
\end{tabular}
\end{adjustbox}
\end{table}

In order to validate that we dealt with liquid rope coiling (LRC) rather than solid rope coiling (SRC) in this study, the effective bending stress (\(B_{\text{eff}}\)) and effective viscosity of the inks were established by using the formulas presented below for prediction of the coiling frequency of liquid and solid materials in the gravitational regime \cite{Habibi2007, Maleki2004, Rahmani2011}. This \(B_{\text{eff}}\) gives an indication of the stress in the filament during extrusion.

\[ \Omega_{G,\text{liquid}} = \left( \frac{gQ^3}{\nu a_1^8} \right)^{1/4} \]

\[ \Omega_{G,\text{solid}} = U \left( \frac{\rho g}{d^2 B} \right)^{1/3} \]

where:
\begin{itemize}
    \item \(\Omega_{G,\text{liquid}}\) is the coiling frequency for liquid materials,
    \item \(\Omega_{G,\text{solid}}\) is the coiling frequency for solid materials,
    \item \(g\) is the gravitational acceleration,
    \item \(Q\) is the volumetric flow rate,
    \item \(\nu\) is the kinematic viscosity,
    \item \(a_1\) is the radius of the filament in the coil,
    \item \(U\) is the printing speed,
    \item \(\rho\) is the density of the material,
    \item \(d\) is the diameter of the nozzle,
    \item \(B\) is the bending stiffness.
\end{itemize}

From Supplementary Table \ref{tab: SI rheo ink properties}, it can be observed that \( B_{\text{eff}} \) of all samples falls within an order of magnitude of $\sim 10^1$ kPa, and the effect of nozzle geometry is reflected by the fact that \( B_{\text{eff}} \) for the cookie ink printed with a 1.0 mm nozzle is around two times greater than that printed with a 1.5 mm nozzle; i.e., the stress is higher for smaller nozzles. The \( \sigma_{\text{yield}} \) of all inks is one order of magnitude smaller than \( B_{\text{eff}} \), indicating that the inks do not behave like an elastic material. The effective kinematic (\( \nu_{\text{eff}} \)) and dynamic (\( \eta_{\text{eff}} \)) viscosities (Supplementary Table \ref{tab: SI rheo ink properties}) are considerably high, being around 3 orders of magnitude higher than the viscosity of honey at room temperature. The inks must, therefore, flow like a very thick viscous fluid. With these approximations, and the aforementioned \( n \) and \( K \) indices, it is concluded that all inks display a liquid behavior during extrusion with sufficient viscosity to exhibit coiling, meaning that the LRC formulas can indeed be used to predict the coiling behavior accurately.

\subsection{Ink stability}

In addition to being able to pass through a nozzle and to exhibit coiling behavior, the ink properties should also allow for a fast stabilization once the ink has been deposited, to ensure that the printed coiled structure does not collapse \cite{Tejada-Ortigoza2022}. This post-printing stability is mostly dependent on the ink’s viscoelastic moduli, \( G' \) and \( G'' \), and loss tangent \( \tan \delta \) under large oscillatory shear (LOAS) ranges, as these indicate how much stress a material can withstand prior to collapsing. From Figure \ref{fig:rheology}d (main text), it can be observed that \( G' \) remains nearly constant in the linear viscoelastic regime (LVR), regardless of ink type, until a critical strain is reached, after which it falls abruptly. This indicates the transition to the nonlinear viscoelastic regime (NLVR), where the material’s network starts to break down, resulting in shear-thinning behavior \cite{Hyun2011}. In the LVR, inks have \( G' \) values greater than the minimum of 300 Pa reported in literature for post-printing stability \cite{Outrequin2023}, so therefore are guaranteed to yield stable prints. Moreover, the \( \tan \delta \) of the inks in the LVR are 0.18, 0.13 and 0.36 for pea, carrot and cookie, respectively. The former two fit within a range of \( 0.052 \leq \tan \delta \leq 0.268 \), which was found to deliver stable prints \cite{Chen2019, Gholamipour-Shirazi2019, Yang2021}. The \( \tan \delta \) of cookie was closer to one, indicating that it is closer to exhibiting flow behavior, and would, therefore, be less self-supporting. However, the previously indicated range was established for food inks with very different compositions, so may not act as an accurate representation for the inks used in this study.

Furthermore, for application purposes, the longevity of ink stability was investigated to establish if the inks could be used for multiple days on end. In Supplementary Figure \ref{fig:SI ink stability}, it can be seen that all inks, except carrot, remain stable for at least 24 hours, with cookie ink remaining stable up to five days. However, all inks have high \(A_w\) values: \( 0.98 \pm 0.00 \), \( 1.00 \pm 0.01 \), and \( 0.84 \pm 0.01 \), for pea, carrot, and cookie, respectively. This indicates that their microbial stability may deteriorate fast and should be monitored closely.

\begin{figure}[!h]
    \centering
    \includegraphics[width=0.9\textwidth]{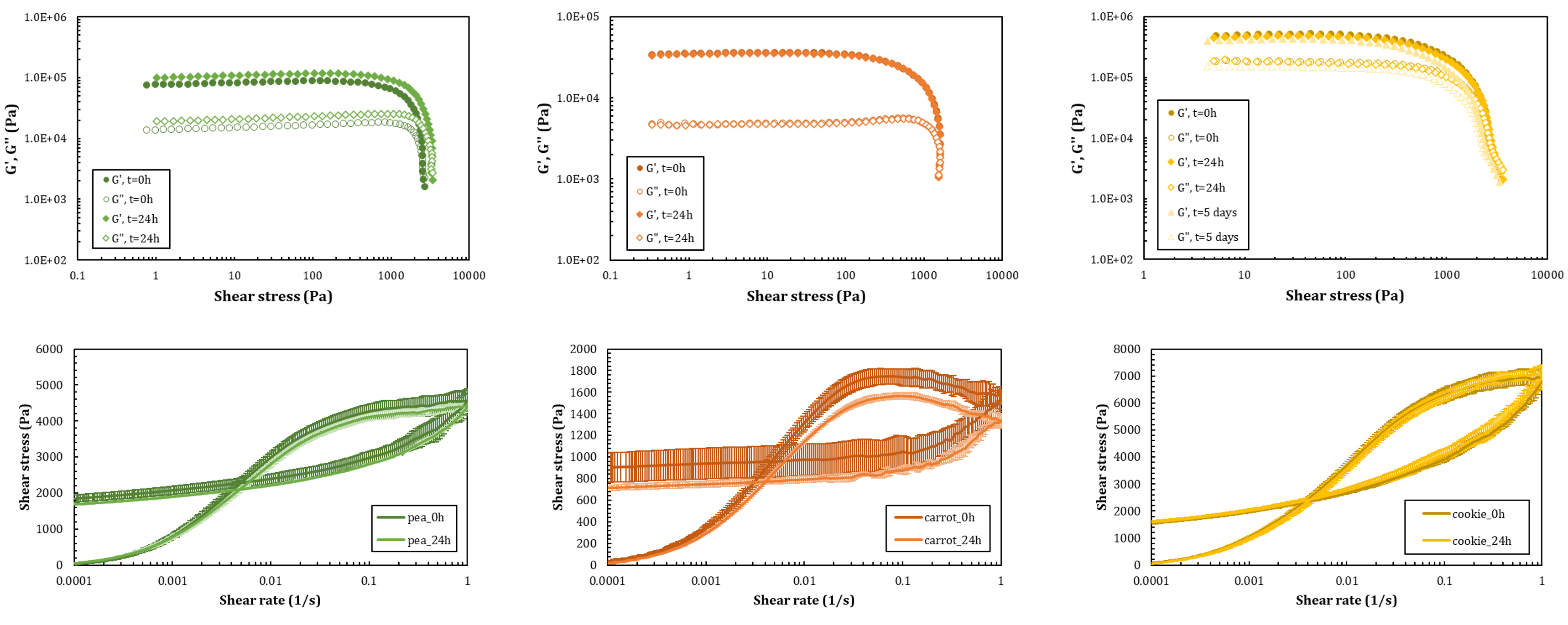}
    \caption{Ink stability over time. Top row: amplitude sweeps of pea (green), carrot (orange) and cookie inks (yellow). Bottom row: steady shear sweeps.}
    \phantomsection
    \label{fig:SI ink stability}
\end{figure}

\subsection{Evaluation of Lissajous plots and nonlinear response upon LAOS}

Additional Lissajous plot analyses and quantifications from the LAOS deformation are able to provide more details on the inks during their transition from the LVR to the NLVR region, which is also what occurs during extrusion from the nozzle during 3DFP. This type of more in-depth rheological analyses through the use of Lissajous plots of food-based inks have only recently started to gain attention by scientists \cite{Shahbazi2021, Yu2022, Yu2022b, Yu2023}. From the elastic Lissajous plots in Supplementary Figure \ref{fig:SI lissajous}a, it can be observed that at 0.01\% strain, all inks exhibit a linear viscoelastic response, characterized by their narrow elliptical shape. As the imposed strain amplitude increases from 1 to 10\%, the ellipses start to widen as a result of \( G' \) decreasing faster than \( G'' \) (Supplementary Figure \ref{fig:SI lissajous}b). Ultimately, the data shifts to a parallelogram shape, implying a shift to a nonlinear viscoelastic response; i.e., from a solid-dominated to a liquid-dominated behavior \cite{Ewoldt2008, Hyun2011}. This is also what occurs during the extrusion of ink from the nozzle. Furthermore, the largest change in shape under large strain amplitudes is observed for cookie and pea inks, which indicates a greater structural breakdown of the inks under large deformation, although complete microscopic fracture did not occur within the strain range tested.

\begin{figure}[!h]
    \centering
    \includegraphics[width=0.9\textwidth]{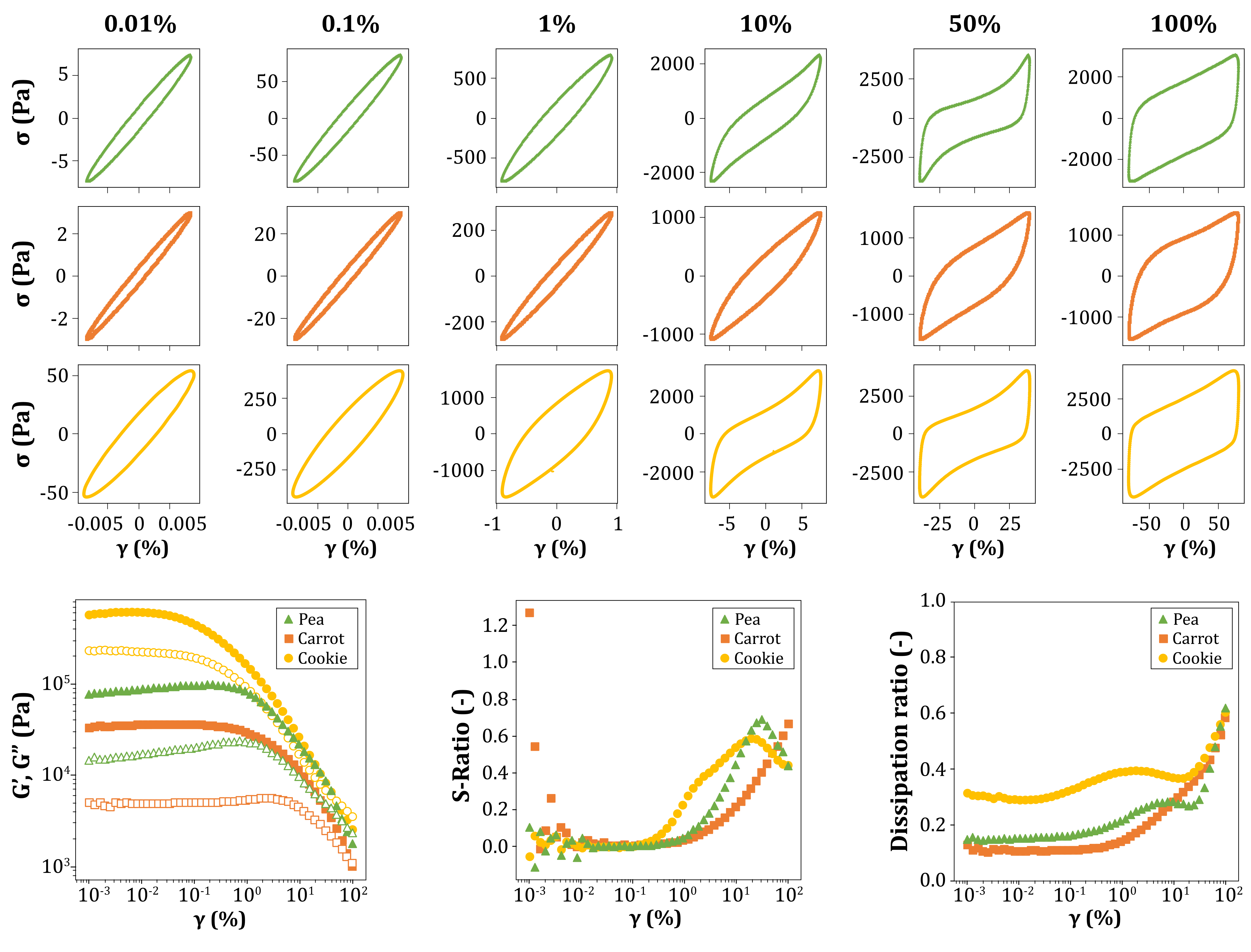}
    \caption{
    (a) The elastic Lissajous plots for pea (green), carrot (orange), and cookie (yellow) inks. 
    (b) Viscoelastic moduli versus shear strain; \( G'\)=solid symbols, \( G''\)=open symbols. 
     (c) Stiffening ratios. 
    (d) Dissipation ratio; \( \sigma \)=stress, \( \gamma \)=strain.}
    \phantomsection
    \label{fig:SI lissajous}
\end{figure}

From the Lissajous plots, further nonlinear LAOS responses can be quantified by using the strain-stiffening (S) ratio and the dissipation ratio, which describe the intracycle strain stiffening as determined from the local Lissajous slopes, and the ratio between the energy dissipated and the energy stored by the material during deformation, respectively \cite{Ewoldt2008, Ewoldt2010}. To the best of our knowledge, these ratios have not been reported in literature for 3DFP inks, although various studies on wheat doughs, tomato puree, and mashed potato are available, which bear some resemblance with the inks in this study \cite{Duvarci2017, Joyner2016, Turksoy2021, Yazar2016}. Supplementary Figure \ref{fig:SI lissajous}c shows that the S-ratios remain close to 0 until an imposed strain of around 1\%, indicating that no strain stiffening occurs. Upon further increase of the strain, the S-ratio increases, indicating intracycle strain stiffening; a phenomenon often reported during the initiation of steady shear thinning in food materials \cite{Duvarci2017, Klost2020, Precha-Atsawanan2018, Turksoy2021, Yazar2016}. The S-ratio of the cookie ink starts deviating from 0 before the other two inks, which can also be observed from the Lissajous curve where it has the most drastic shape change of all inks between 1 and 10\% strain. Finally, Supplementary Figure \ref{fig:SI lissajous}d demonstrates the increasing energy dissipation of the different inks at higher strains, as also indicated by the growth in the enclosed area of the Lissajous curves. The cookie ink acts the most dissipative and so the least elastic, also corroborated by it having the largest enclosed Lissajous area. Carrot, on the other hand, exhibits the weakest strength (low \( G' \)), whereas it is the most elastic due to its dissipation ratio being close to zero.

The differences in rheological behavior among the different inks can be ascribed to their composition. It is commonplace to link this difference to the microstructure of the materials; however, as was elaborated upon in the previous section, this was challenging for the inks used in this study. The only characteristic of the inks that is known is their moisture content (MC). The carrot ink has the highest MC of \( 87.90 \pm 0.08 \% \) wb, which resulted in it having the weakest network due to the high presence of water, indicated by the low \( G' \). The MC of pea is \( 53.10 \pm 0.03 \% \) wb, followed by cookie with the lowest MC of \( 15.81 \pm 0.15 \% \) wb. This low MC of the cookie ink, together with the fact that it comprises multiple ingredients, including fat, starch, and gluten, causes it to have the strongest network and thus the highest \( G' \). All things considered, these results prove that all inks used in this study can easily be extruded through the nozzle, provided that the printer motor can exert sufficient shear force \cite{Shahbazi2021}, and that all are capable of exhibiting coiling behavior.

\section{Further information on coiling}

\subsection{Changing the flow rate through modification of EM and U\textsubscript{xy}}

The extrusion multiplier (EM) and the movement speed of the printer head (U\textsubscript{xy}) can be individually set during 3D printing, both of which affect the overall flow rate (Q). Higher speeds are often preferred, although they may result in reduced precision and compromised print quality. Slower speeds, on the other hand, enable more precise control over material placement, leading to better printing resolution. The intricate balance between EM and U\textsubscript{xy} plays a vital role in determining the Q and the quality of the print, so finding the optimal balance between the two is crucial to achieve prints with the desired characteristics. During conventional extrusion-based 3DFP, small variations in EM and U\textsubscript{xy} (e.g., a 0.1 change in EM) may already have drastic effects on the printed structure. This effect is less pronounced during 3DFP processes that make use of the LRC effect, as the printed layers do not get deposited on top of each other directly.  

By individually changing the EM and U\textsubscript{xy}, we visualized their individual effects on 3DFP and coiling specifically. Moving horizontally in Supplementary Figures \ref{fig:SI EM and Uxy}a-c, it can be observed that when keeping EM the same, differences among prints are minimal; confirming that printing speed does not affect the result much. The EM does, however, significantly affect the print. The coiling frequencies corresponding to the different combinations of printing parameters increase linearly with the flow rates presented in Supplementary Figure \ref{fig:SI EM and Uxy}d for each combination of EM and U\textsubscript{xy}, so increase diagonally from the top left to bottom right. In this study, coiling up to 2000 mm/min was successfully tested, but this is dependent on the printer and its motor capacity. Furthermore, an interesting observation is that prints with equal flow rates do not yield similar prints, as can be seen with e.g., those printed with \(Q=24.48 \, \text{mm}^3/\text{s}\).

\begin{figure}[!h]
    \centering
    \includegraphics[width=0.9\textwidth]{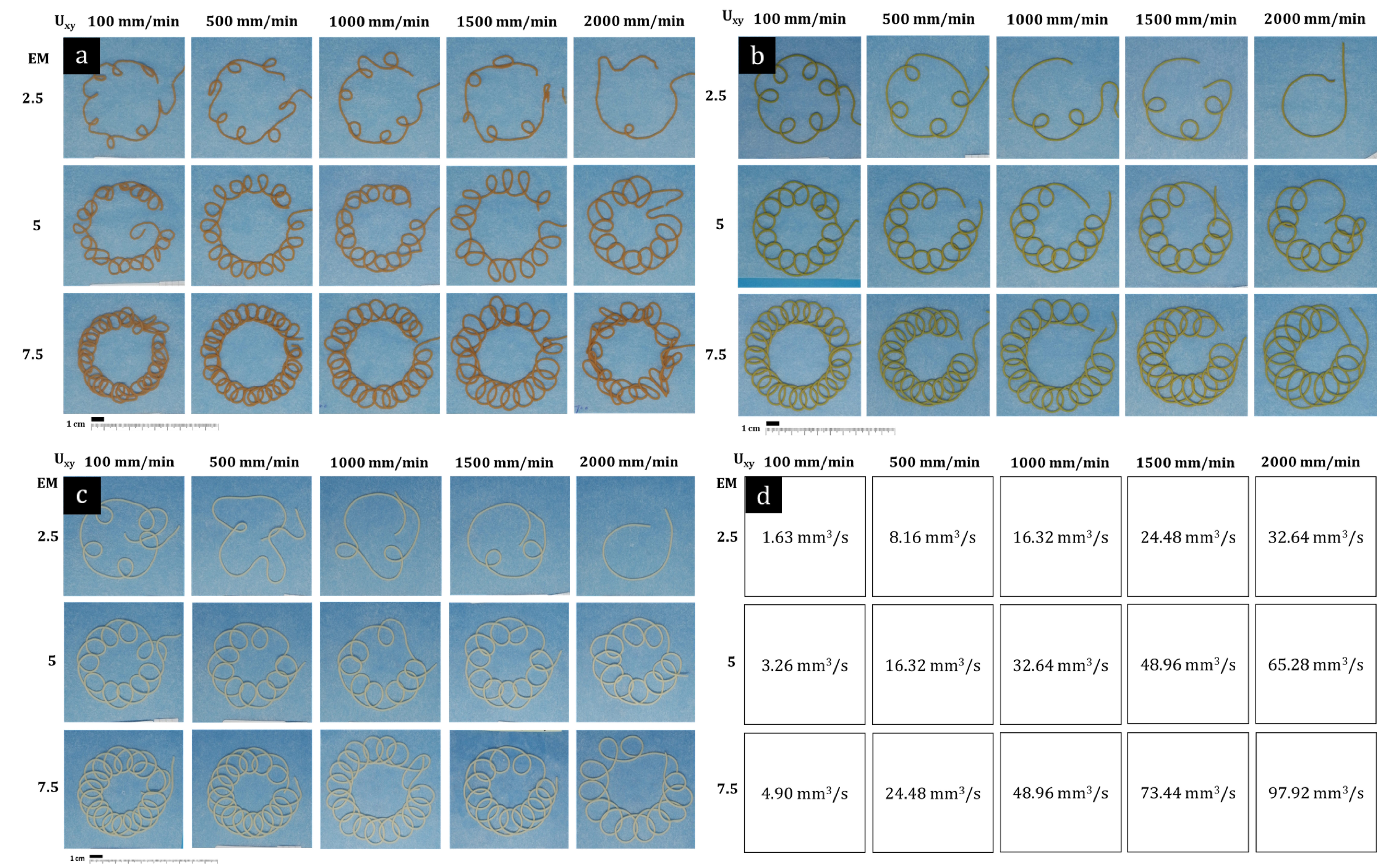}
    \caption{Effects of different printing speed and extrusion multiplier ratios on the coiling behavior of 
    (a) pea, 
    (b) carrot, and 
    (c) cookie inks printed with a 1.5 mm nozzle at a height of 4 cm. 
    (d) Flow rates corresponding to each combination of EM and \(U_{xy}\).}
    \phantomsection
    \label{fig:SI EM and Uxy}
\end{figure}

\subsection{The effects of printing parameters on coiling behavior}

\textbf{Moving belt versus moving nozzle}

\begin{figure}[!h]
    \centering
    \includegraphics[width=0.9\textwidth]{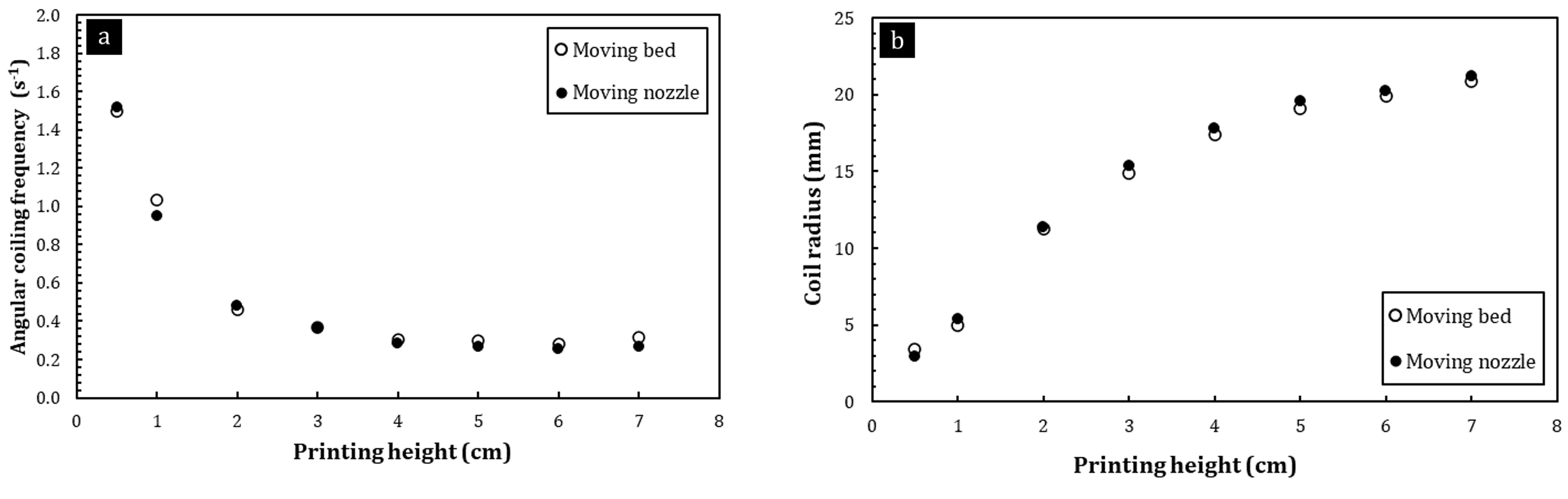}
    \caption{Differences in coiling behavior initiated by moving the bed or moving the nozzle in terms of 
    (a) coiling frequency and 
    (b) coil radius as a function of printing height.
    }
    \phantomsection
    \label{fig:SI coiling initiation}
\end{figure}

The experiments in this study were executed on a printer with a moving (x, y, z) nozzle and a stationary platform (TNO ‘SUPREME’ printer). In literature, studies on dynamic coiling are often performed with a setup that involves a moving belt while the orifice from which the material gets deposited remains stationary. It was assumed that both methods induce similar coiling and sewing behaviors, but to confirm this, we repeated the printing height experiments with another 3D food printer that used a system in which the nozzle moved in the x,z-directions and the platform moved in the y-direction (TNO ‘PRIME’ printer). From Supplementary Figure \ref{fig:SI coiling initiation}, it can be observed that both coil frequency and radii are nearly identical for both printing methods, which confirms that both printing methods can be used interchangeably to print coiled structures.

\textbf{Coiling behavior along straight and circular paths}

\begin{figure}[!h]
    \centering
    \includegraphics[width=0.9\textwidth]{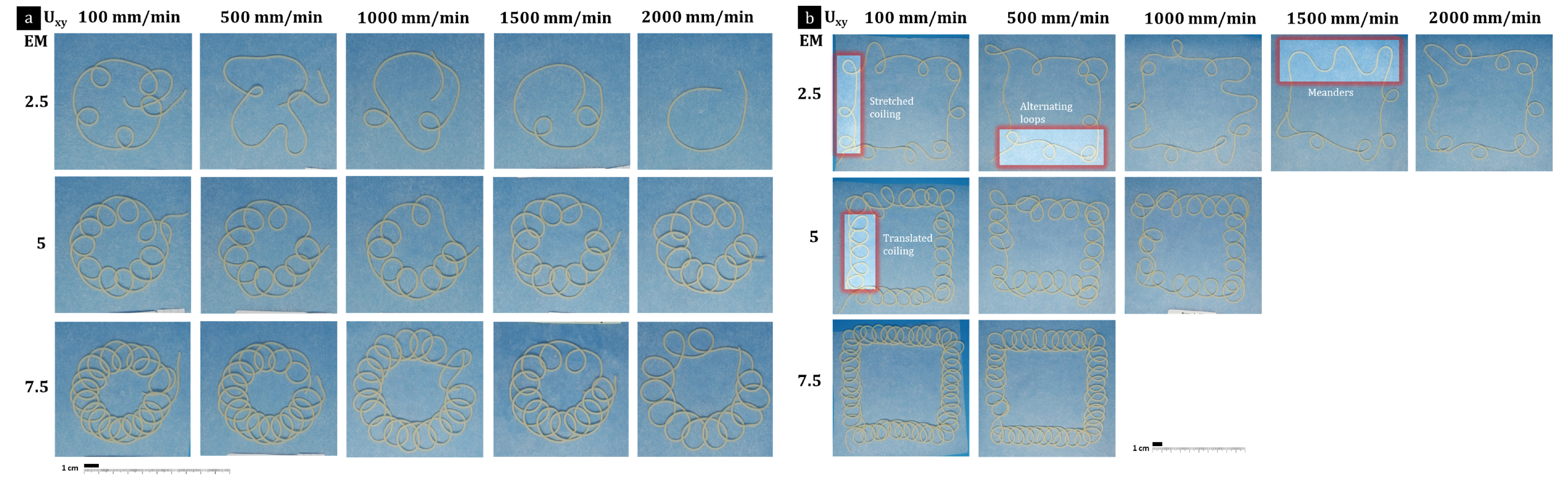}
    \caption{Coiling under similar conditions along 
    (a) circular and 
    (b) square paths with diameters of 6 and 10 cm, respectively. The highlighted areas depict specific categories of sewing patterns.
    }
    \phantomsection
    \label{fig:SI path shape}
\end{figure}

During the experiments, it was observed that the coils would often turn inwards (Supplementary Figure \ref{fig:SI path shape}a), whereas in theory, the direction of coiling should occur randomly. This phenomenon led to the hypothesis that centrifugal forces affected coiling in circular paths, favoring inward movement. So, in order to investigate the effect of printing path on coiling, experiments with square printing paths were executed, as these would eliminate the influence of said forces. Supplementary Figure \ref{fig:SI path shape}b illustrates that the coiling of square paths does, indeed, follow a seemingly more random starting direction as compared to circular ones, although the sample size should be increased for further clarification. 

\begin{figure}[!h]
    \centering
    \includegraphics[width=0.9\textwidth]{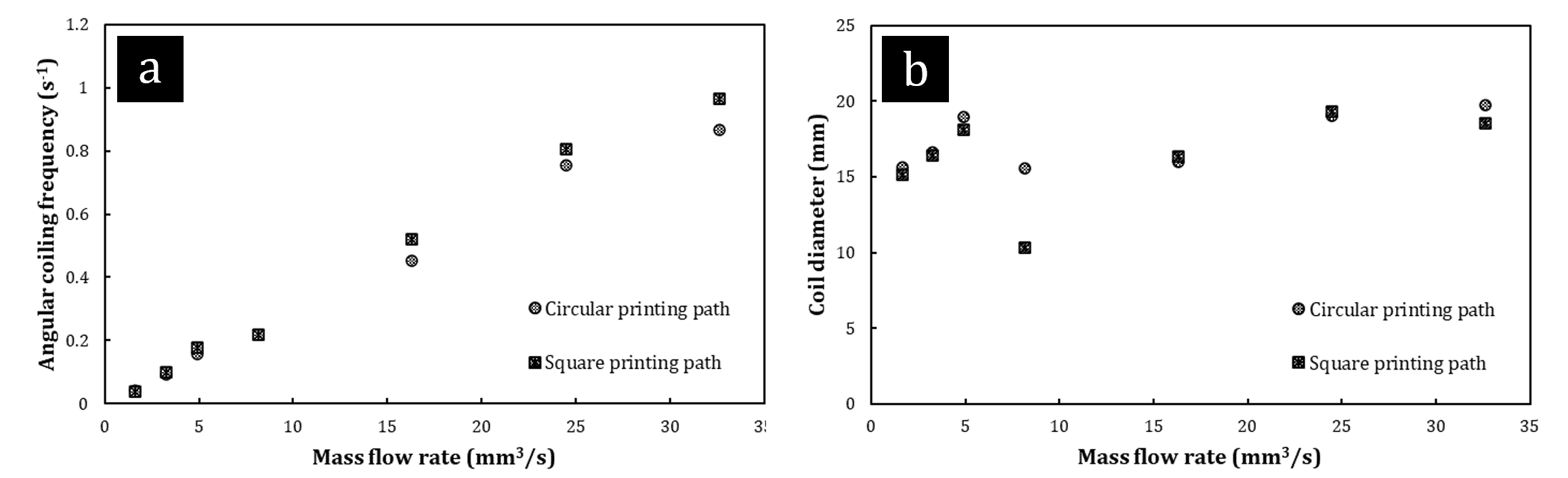}
    \caption{
    (a) Angular coiling frequency and 
    (b) coil diameter as a function of mass flow rate for circular and square printing paths.
    }
    \phantomsection
    \label{fig:SI coiling parameters path}
\end{figure}

Furthermore, when using straight printing paths, a greater variety of coiling patterns could be achieved, albeit only at low EM values (Supplementary Figure \ref{fig:SI path shape}b). However, all patterns except translated coiling were observed to not be very stable; they tended to only last a few repetitions before changing into another pattern. This pattern formation is dependent on the ratio between the injection (\(U\)) and the nozzle (\(U_{xy}\)) speeds. For the EM 2.5 row, \(U:U_{xy}\) was 0.55, for EM 5 it was 1.11, and for EM 7.5 it was 1.66. According to theoretical frameworks established by \cite{Chiu2006, Jawed2014, Morris2008}, there should not be any coiling nor stable material deposition at EM 2.5 due to \(U << U_{xy}\) applying. However, our study contradicted this expectation. Only at EM 2.5 and \(U_{xy} = 2000 \) mm/min, some slippage occurred at the circular printing path. In the middle row, \(U \approx U_{xy}\), so in theory, a steady line should have been formed – which again, did not occur. Pattern formation should only have started occurring when \(U >> U_{xy}\), so in the bottom row. The reason why the sewing in our study deviated from trends described in literature remains unknown, although the difference between \(U\) and \(U_{xy}\) may not have been high enough. Another possible explanation may be that the inks used in this study had a viscosity that was multiple orders of magnitudes higher than those of inks used in previous studies. This increased amount of energy stored in the ink may have affected its coiling behavior. Nevertheless, translated coiling was found to be the most prevalent sewing pattern, which was the primary focus in this study. Moreover, no significant difference was observed between \(R\) and \(\Omega\) of the coils formed along either square or circular paths (Supplementary Figure \ref{fig:SI coiling parameters path}), leading to the belief that, besides the coiling direction, the shape of the path has only minor effects on the coils themselves.

\subsection{Printing parameters used for the creation of the coiled structures}

See Supplementary Figure \ref{fig:SI printing parameters}

\begin{figure}[!h]
    \centering
    \includegraphics[width=0.9\textwidth]{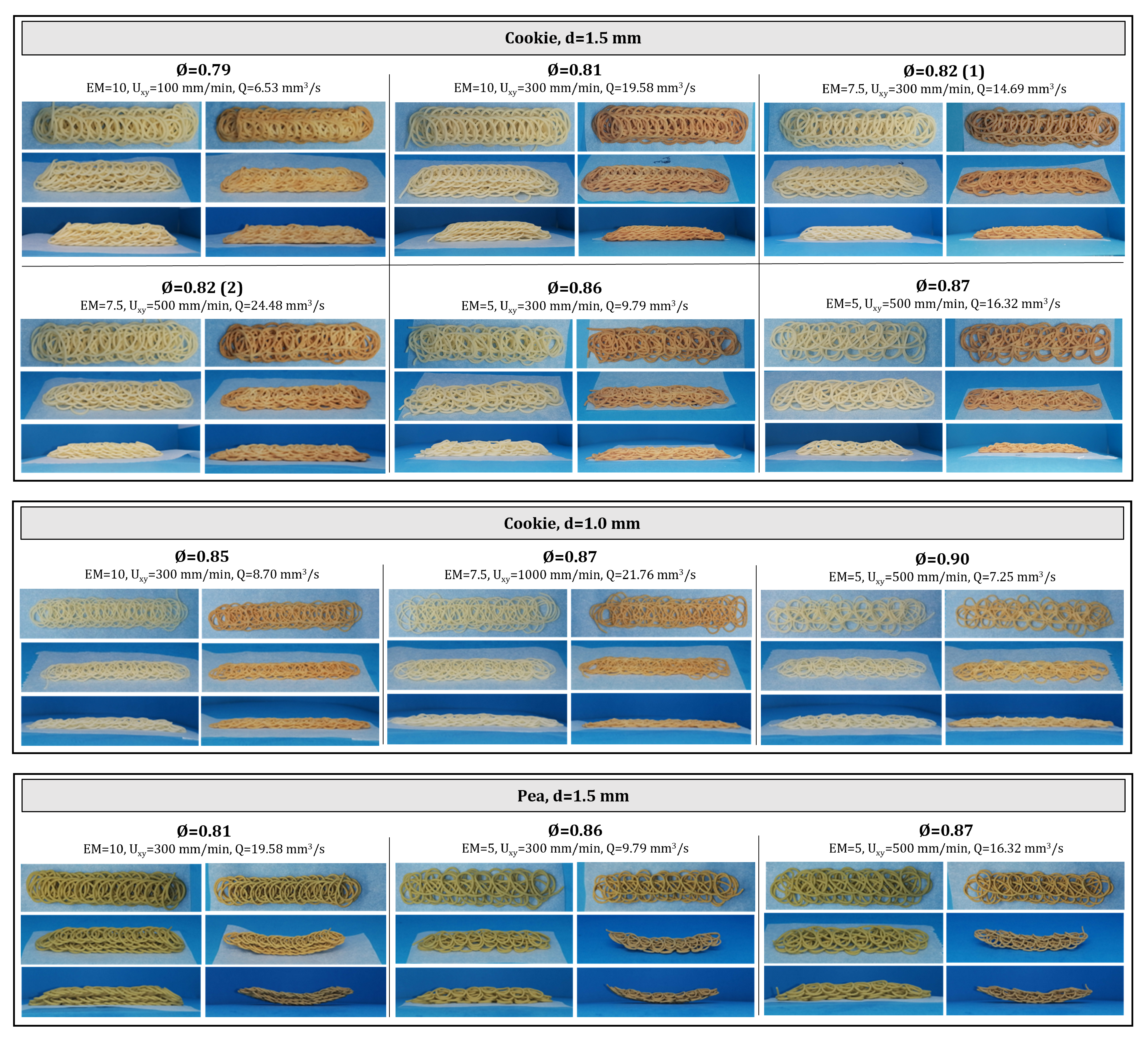}
    \phantomsection
    \label{fig:SI printing parameters}
    \caption{Printing parameters used to create coiled structures (left: unbaked, right: air-fried) with varying degrees of porosity.
    }
\end{figure}

\subsection{Porosity of the coiled structures}
See Supplementary Table \ref{tab:SI porosity}

\begin{table}[!htb]
	\centering
	\caption{External porosities (expressed as air fraction) of the coiled structures, calculated by using the sample's square outer volume ($ \varnothing_{\text{macro}} $) and its exact outer volume determined by X-ray computed tomography ($ \varnothing_{\text{XRT,macro}} $), and the internal porosity within the filament strands ($ \varnothing_{\text{XRT,micro}} $).}
	\phantomsection
	\label{tab:SI porosity}
	\begin{adjustbox}{width=\textwidth, margin=0pt 0pt 0pt 0cm} 
		\begin{tabular}{lcccc}
			\toprule
			\textbf{Ink Type} & $\mathbf{\varnothing_{\text{external}}}$ & $\mathbf{\varnothing_{\text{actual}}}$ & $\mathbf{\varnothing_{\text{external}} - \varnothing_{\text{actual}}}$ & $\mathbf{\varnothing_{\text{bulk}}}$ \\
			\midrule
			\multirow{3}{*}{Cookie, $d = 1.5 \, \text{mm}$} & 0.81 & 0.54 & 0.27 & 0.17 \\
			& 0.86 & 0.61 & 0.25 & 0.15 \\
			& 0.87 & 0.60 & 0.27 & 0.15 \\
			\midrule
			Cookie, $d = 1.0 \, \text{mm}$ & 0.87 & 0.68 & 0.19 & 0.13 \\
			\midrule
			Pea, $d = 1.5 \, \text{mm}$ & 0.81 & 0.72 & 0.09 & 0.15 \\
			\bottomrule
		\end{tabular}
	\end{adjustbox}
\end{table}

\section{Texture analysis}

\subsection{Fracture behavior of bulk materials} 

\begin{figure}[!h]
    \centering
    \includegraphics[width=0.9\textwidth]{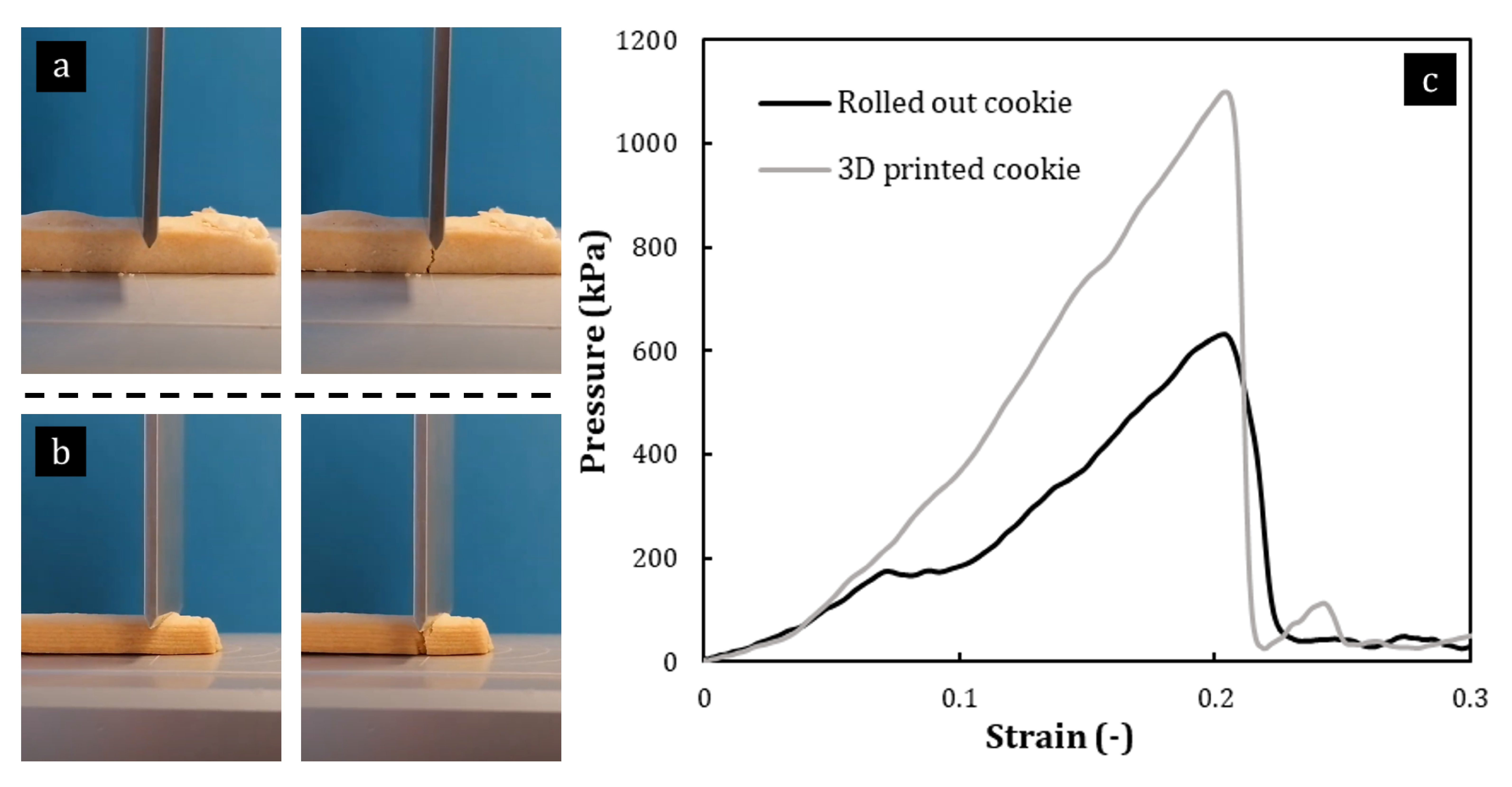}
    \caption{Cutting test on baked cookies that were made by (a) traditional rolling out of the dough and by 
    (b) 3D printing a structure with 100\% infill. 
    (c) Fracture behavior of both cookies. 
    }
    \phantomsection
    \label{fig:SI bulk fracture}
\end{figure}

In order to compare the fracture behavior of the coiled structure to that of a ‘conventional’ cookie, we subjected rolled-out cookies and solid 3D printed cookies to the same series of cutting tests. For this, the same cookie dough that was also used as 3D printing ink was used. The 3D printed structure had 100\% infill, and both the printed and rolled-out cookies had dimensions similar to those of the coiled structure, namely approximately 27x110x8 mm (WxLxH). Supplementary Figure \ref{fig:SI bulk fracture} shows one of five measurements taken. Although the results are similar to those of cookies reported in literature, it should be noted that the recipe used in this study does not contain any leavening agent such as baking powder. Conventional cookies usually contain ingredients like this, which may induce a slightly different fracture behavior than what is reported here.

\subsection{Fracture of a single filament}

\begin{figure}[!h]
    \centering
    \includegraphics[width=0.9\textwidth]{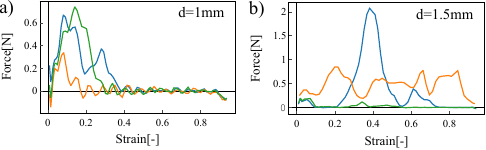}
    \caption{
    Cutting force response of cookie filaments with diameters of (a) 1 mm and (b) 1.5 mm, shown in three repetitions.
    }
    \phantomsection
    \label{fig:SI filament fracture}
\end{figure}

Cutting a single filament allows us to quantify the minimum meaningful fracture force. The highest fracture force response of the filament is equivalent to the minimum meaningful fracture in our coiled structures. Supplementary Figure \ref{fig:SI filament fracture}, shows the applied force as a function of the applied strain for cookie filaments.

\subsection{Initial stiffness of coiled pea structures} 
The initial stiffness for different porosities is presented in Supplementary Figure \ref{fig:SI Y_modulus}. 
The initial stiffness shows considerable variation among the samples; however, its order of magnitude consistently falls within the range of $10^6 \text{Pa}$ during compression and cutting tests.

\begin{figure}[!h]
    \centering
    \includegraphics[width=0.6\textwidth]{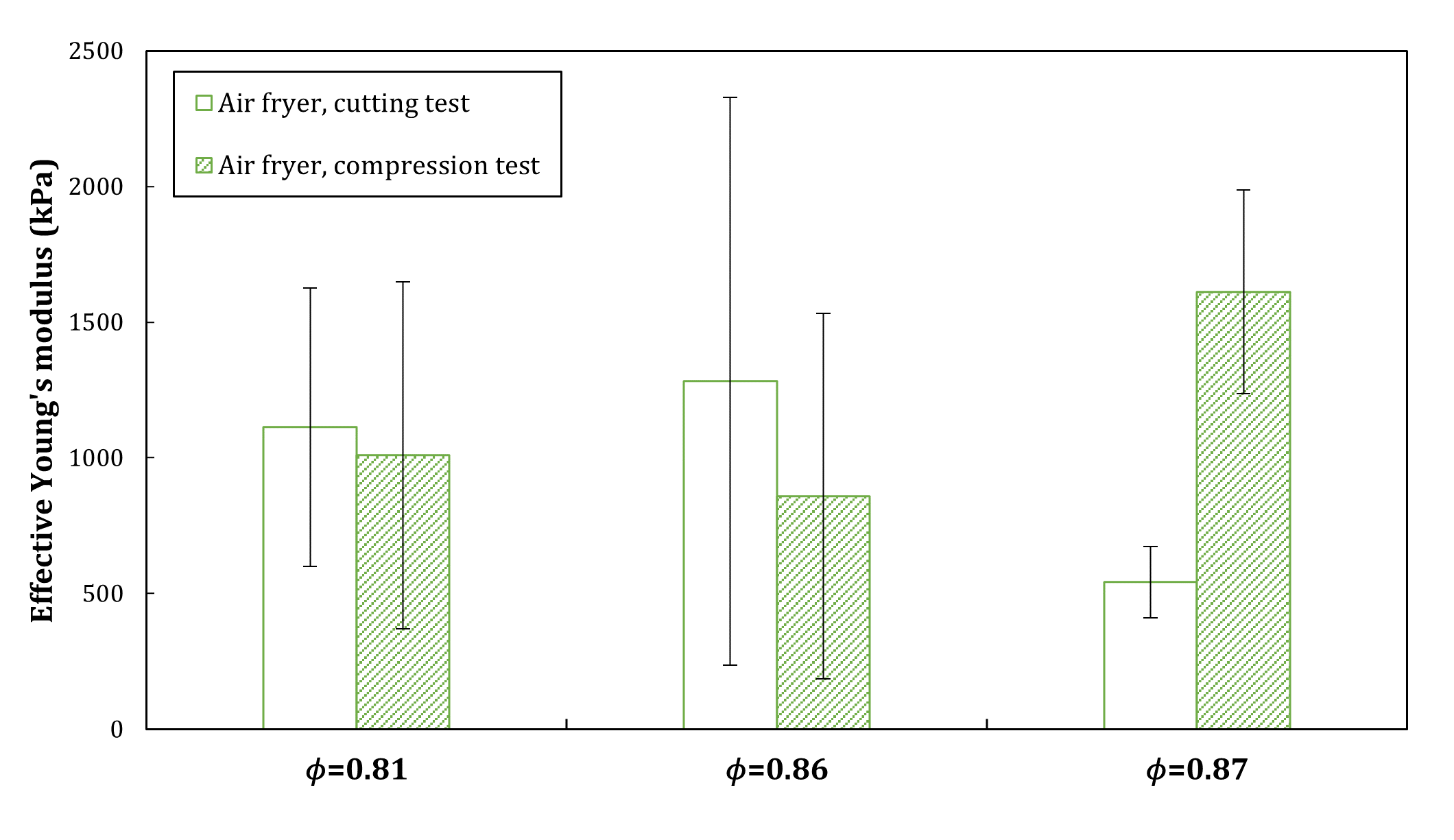}
    \caption{Initial stiffness of air fried pea structures with increasing porosity, printed with a 1.5 mm nozzle.
    }
    \phantomsection
    \label{fig:SI Y_modulus}
\end{figure}

\end{document}